\pdfoutput=1
\documentclass[sigconf,bookmarks=false]{acmart}

\makeatletter                   
\def\mdseries@tt{m}             
\makeatother                    
\usepackage[plain]{fancyref}
\usepackage[draft=true]{minted} 
\usepackage{color}
\usepackage{hyperref}           
\hypersetup{
    colorlinks=true,
    linkcolor=blue,
    filecolor=red,      
    urlcolor=magenta,
    breaklinks=true,            
}
\usepackage{breakurl}           

\usepackage{multirow}
\usepackage{amsmath}
\usepackage{amsfonts}
\usepackage{framed}
\usepackage[skins]{tcolorbox}
\usepackage{tikz}
\usetikzlibrary{backgrounds}
\usepackage[framemethod=TikZ]{mdframed}
\usepackage{listings}
\usepackage{color}
\usepackage{algorithm}
\usepackage[noend]{algpseudocode}
\usetikzlibrary{shadows}
\usepackage{flushend}
\usepackage{xspace}
\usepackage{minted}
\usepackage{pbox}
\usepackage{balance}
\usepackage{soul}

\usepackage{balance}

\setminted{style=borland}

\definecolor{dkgreen}{rgb}{0,0.6,0}
\definecolor{gray}{rgb}{0.5,0.5,0.5}
\definecolor{mauve}{rgb}{0.58,0,0.82}
\definecolor{codegreen}{rgb}{0,0.6,0}
\definecolor{codegray}{rgb}{0.5,0.5,0.5}
\definecolor{codepurple}{rgb}{0.58,0,0.82}
\definecolor{backcolour}{rgb}{0.95,0.95,0.92}

\newcommand{\etal}{et al. }

\newcommand{\figref}[1]{Figure~\ref{fig:#1}}
\newcommand{\tref}[1]{Table~\ref{tab:#1}}
\newcommand{\aref}[1]{Algorithm~\ref{alg:#1}}
\newcommand{\sect}[1]{Section~\ref{sect:#1}}

\newcommand{\tbi}[1]{\textbf{\textit{#1}}}

\newcounter{resQues}
\newcommand{\theResQues}{\arabic{resQues}}

\newenvironment{resQues}[1][]{%
 
    \ifstrempty{#1}%
    {
    \refstepcounter{resQues}
    \mdfsetup{%
        frametitle={%
            \tikz[baseline=(current bounding box.east),outer sep=5pt]
            \node[anchor=east,rectangle,fill=blue!20]
            {\strut Research Question~\theResQues};}
        }%
    }{\mdfsetup{%
        frametitle={%
            \tikz[baseline=(current bounding box.east),outer sep=5pt]
            \node[anchor=east,rectangle,fill=blue!20]
            {\strut Research Question ~#1};}%
        }%
    }%
    \mdfsetup{%
        innertopmargin=10pt,linecolor=blue!20,%
        linewidth=1pt,topline=true,%
        frametitleaboveskip=\dimexpr-\ht\strutbox\relax%
    }
\vspace{4mm}\begin{mdframed}[]\vspace{-4mm}\relax}{%
\end{mdframed}}

\newcounter{todoCounter}
\newenvironment{TODOEnv}[1][]{
    \refstepcounter{todoCounter}
    \noindent \textbf{TODO~\arabic{todoCounter}: ~#1}
}

\newcommand{\added}[1]{\textcolor{black}{{#1}}}

\newcommand{\baseline}{HitoshiIO}
\newcommand{\approach}{SLACC\xspace}
\newcommand{\longapproach}{Simion-based Language-Agnostic Code-Clone detection\xspace}

\newcommand{\PERSON}{Avery\xspace}

\usepackage{tabularx}
\usepackage{booktabs}
\usepackage{array}
\usepackage{multirow}

\usepackage{xcolor, colortbl}
\definecolor{my-blue}{RGB}{83,87,118}

\usepackage{arydshln}
\usepackage{siunitx}
\makeatletter
\def\adl@drawiv#1#2#3{%
        \hskip.5\tabcolsep
        \xleaders#3{#2.5\@tempdimb #1{1}#2.5\@tempdimb}%
                #2\z@ plus1fil minus1fil\relax
        \hskip.5\tabcolsep}
\newcommand{\cdashlinelr}[1]{%
  \noalign{\vskip\aboverulesep
           \global\let\@dashdrawstore\adl@draw
           \global\let\adl@draw\adl@drawiv}
  \cdashline{#1}
  \noalign{\global\let\adl@draw\@dashdrawstore
           \vskip\belowrulesep}}
\makeatother

\AtBeginDocument{%
  \providecommand\BibTeX{{%
    \normalfont B\kern-0.5em{\scshape i\kern-0.25em b}\kern-0.8em\TeX}}}

\copyrightyear{2020}
\acmYear{2020}
\setcopyright{acmcopyright}
\acmConference[ICSE '20]{42nd International Conference on Software Engineering}{May 23--29, 2020}{Seoul, Republic of Korea}
\acmBooktitle{42nd International Conference on Software Engineering (ICSE '20), May 23--29, 2020, Seoul, Republic of Korea}
\acmPrice{15.00}
\acmDOI{10.1145/3377811.3380407}
\acmISBN{978-1-4503-7121-6/20/05}

\begin{document}
\sloppy
%
\title{\approach{}: Simion-based Language Agnostic Code Clones}

\author{George Mathew, Chris Parnin, Kathryn T Stolee}
\affiliation{%
  \institution{North Carolina State University}
}
\email{{george2, cjparnin, ktstolee}@ncsu.edu}

\renewcommand{\shortauthors}{Mathew \etal}

\begin{abstract}

Successful cross-language clone detection could enable researchers and developers to create robust language migration tools, facilitate learning additional programming languages once one is mastered, and promote reuse of code snippets over a broader codebase. 
However, identifying cross-language clones presents special challenges to the clone detection problem.
A lack of common underlying representation between arbitrary languages means detecting clones requires one of the following solutions: 1) a static analysis framework replicated across each targeted language with annotations matching language features across all languages,  or 2) a dynamic analysis framework that detects clones based on runtime behavior. 

In this work, we demonstrate the feasibility of the latter solution, a dynamic analysis approach \added{called \approach{}} for cross-language clone detection. 
Like prior clone detection techniques, we use input/output behavior to match clones, though we overcome limitations of prior work by amplifying the number of inputs and covering more data types; and as a result, achieve better clusters than prior attempts.
\added{Since clusters are generated based on input/output behavior, SLACC supports cross-language clone detection.}
As an added challenge, we target a static typed language, Java, and a dynamic typed language, Python.
Compared to \baseline{}, a recent clone detection tool for Java, \approach{} retrieves $6$ times as many clusters and has higher precision ($86.7\%$ vs. $30.7\%$).

This is the first work to perform clone detection for dynamic typed languages (precision = $87.3\%$) and the first to perform clone detection across languages that lack a common underlying representation (precision = $94.1\%$). It provides a first step towards the larger goal of scalable language migration tools. 

\end{abstract}



\maketitle



%

\section{Introduction}
\label{sect:intro}

Modern programmers typically work on systems built with a cocktail of multiple programming languages~\cite{fjeldberg2008:polyglot}. 
A recent survey found that professional software developers have a mean of seven different programming languages in their industrial software projects~\cite{Mayer2017} and open-source software projects frequently have between 2--5 programming languages~\cite{Tomassetti:2014, Mayer:2015}. Programmers are also expected to continue learning multiple programming languages on a daily basis. To learn a new programming language, studies have shown that programmers attempt to use a \emph{cross-language} learning strategy by reusing knowledge from a previously known language~\cite{quanfeng,  Scholtz:subsequent, Shrestha}. This means programmers often need the ability to relate code snippets across multiple programming languages.

Traditional clone detection often works with only a single programming language, meaning that typical applications and tools are not applicable to modern programming systems and contexts. These applications include bug detection in ported software~\cite{Ray:2013}, maintaining quality through refactoring~\cite{Yue:2018}, and protecting the security of products~\cite{walenstein2007software}. For example, security teams at Microsoft use clone detection to scan for other instances of vulnerable code that might be present in any production software~\cite{Dang:2017}.
In short, there is a need to extend clone detection to work in cross-language contexts, but limited support exists for them.

This paper presents \textbf{S}imion-based \textbf{L}anguage-\textbf{A}gnostic \textbf{C}ode \linebreak \textbf{C}lone detection technique (\approach{}), a \textbf{cross-language} semantic clone detection technique based on code behavior. Our technique can match whole and partial methods or functions. It works in both static and dynamic languages. It does not require annotations or manual effort such as seeding test inputs. Critically, unlike any other clone detection technique, we are able to detect semantically similar code across multiple programming languages and type systems (e.g., Python and Java).

\approach{} finds semantic clones by comparing the input/output (IO) relationship of snippets, called \tbi{simions} (short for \textbf{sim}ilar \textbf{i}nput \textbf{o}utput functio\textbf{ns}), in line with prior work~\cite{jiang2009automatic, su2016identifying}. 
\added{\approach{} segments a target code repository into smaller executable functions. Arguments for the functions are generated using a custom input generator inspired by grey-box testing and multi-modal distribution. Functions are executed on the generated arguments and subsequently clustered based on the generated arguments and corresponding return values. The similarity measure for clustering is based on the IO behavior of code snippets and is independent of their syntactic features. Hence, \approach{}  generates cross-language clusters with code snippets from different programming languages.}
To validate our technique, using a single, static typed language, we perform an empirical study with 19,188 Java functions derived from Google Code Jam (GCJ)~\cite{googleCodeJam} submissions and demonstrate that \approach{} identifies 6x more clones and with higher precision (86.7\% vs. 30.7\%) compared to \baseline{}~\cite{su2016identifying}, a state-of-the-art code semantic clone detection technique.
Using a single, dynamic typed language, we perform a study with 17,215 Python functions derived from GCJ and find that \approach{} can identify true behavioral clones with 87.3\% precision. 
For cross-language clones,  \approach{} finds 32 clusters with both Python and Java functions, demonstrating that detection of code clones does not depend on a common type system. 

In summary, this paper makes the following contributions:
\begin{itemize}
\item For single-language static typed clone detection, an empirical validation demonstrating \approach{} can be used to identify 6x more and better code clones clusters than the state-of-the-art code-clone detection technique \baseline{}.
\item The first exploration of clone detection for a dynamic-typed language and demonstrated feasibility in Python with precision  of 87.3\%.
\item The first exploration of cross-language clone detection when the languages lack an underlying representation; \approach{} is successful in identifying cross-language clone clusters between Python and Java with 94.1\% precision.
\item An \added{open-source} tool for detection of semantic code clones between different programming languages.
\end{itemize}


\section{Motivation}
\label{sect:motivation}

\PERSON is preparing for a technical interview and was given a few practice coding challenges~\cite{10.7717/peerj-cs.173} to work on. \PERSON is more comfortable writing code in Java during an interview setting but is worried because the company exclusively codes in Python. 
As practice for the interview, \PERSON wants to code with Python. 
First, \PERSON decides to write the code in Java to understand the solution, and then translate those solutions into Python code.

One of the practice questions asks the coder to interleave the results of two arrays. \PERSON quickly writes this solution in Java:

\begin{minted}[xleftmargin=2em,linenos,firstnumber=1,fontsize=\small]{Java}
public String interleave(int[] a, int[] b) {
  String result = "";
  int i = 0;
  for( i = 0; i < a.length && i < b.length; i++ ) {
    result += a[i];
    result += b[i];
  }
  int[] remaining = a.length < b.length ? b : a;
  for( int j = i; j < remaining.length; j++ ) {
    result += remaining[j];
  }
  return result;
}
\end{minted}

While one approach is to directly translate the code into Python, \PERSON wonders if there are other ways to take advantage of idioms and capabilities in Python. After spending a few hours searching Stack Overflow~\cite{stackoverflow} and GitHub Gists~\cite{gists}, Avery finds a few code snippets that seem to do the same thing.

The first one seems a bit too complex and relies on another dependency.

\begin{minted}[xleftmargin=2em,linenos,firstnumber=1,fontsize=\small]{Python}
def fancy_interleave(l1, l2):
    from itertools import chain
    return "".join([str(x) 
            for x in chain.from_iterable(zip(l1, l2))])
\end{minted}

This other solution is similar to the Java solution, but is using something new, a \mintinline{Python}{zip} function. Avery is excited to learn some new Python  tricks!

\begin{minted}[xleftmargin=2em,linenos,firstnumber=1,fontsize=\small]{Python}
def problem2(l1, l2):
    result = ""
    for (e1, e2) in zip(l1, l2):
        result += str(e1)
        result += str(e2)
    return result
\end{minted}

\PERSON found the strategy of writing code in Java and translating that code into Python helpful. However, the process of manually searching and translating the code between languages was time-consuming. \PERSON's unfamiliarity with Python made it difficult to verify whether these snippets were \emph{truly} the same. 

At the interview, \PERSON was relieved to be asked to solve the same \emph{interleave} problem from the practice set! 
However, while coding up a solution in Python, the interviewer asked, \emph{does this handle interleaving uneven lists?} The original Java-based solution handled this case, but the Python translation did not. 
Because searching for code took so long, \PERSON never had the chance to fully verify that the Python solution worked the same as the Java solution. 
\PERSON's assumption  that the new \mintinline{Python}{zip} function would work on uneven lists was wrong! 
Had there been  a better way for \PERSON to find semantically related snippets in other programming languages, this issue may have been avoided. 

In this work, we introduce \approach{}, which could detect that these functions are not equivalent. 
From a corpus of code, it could instead find this semantically identical snippet---just one of many applications enabled by cross-language clone detection:
\begin{minted}[xleftmargin=2em,linenos,firstnumber=1,fontsize=\small]{Python}
def valid_interleave1(l1, l2):
    result = ""
    a1, a2 = len(l1), len(l2)
    for i in range(max(a1, a2)):
        if i < a1:
            result += str(list1[i])
        if i < a2:
            result += str(list2[i])
    return result
\end{minted}

\section{\longapproach{}}
\label{sect:thisWork}
Code clones can be broadly classified into four types~\cite{roy2009comparison} as described in \tref{types}. Types I, II and III represent syntactic code clones where similarity between code is estimated with respect to the structure of the code. On the other hand, type-IV indicates functional similarity. Syntactic code clone detection techniques are impractical for cross-language code clone detection as it would require an explicit mapping between the syntax of the languages. This is feasible for syntactically similar languages like Java and C\# ~\cite{java2Csharp} but much harder for different languages like Java and Python. On the other hand semantic approaches for cross-language code detection~\cite{nguyen2017exploring} rely on large number of training examples between the languages and was yet again tested on similar programming languages. 

\begin{table}[b]
    \centering
    \caption{Types of code clones. Types I, II and III are syntactic while type IV are semantic or behavioral clones~\cite{roy2009comparison}}
    \vspace{-6pt}
    \begin{tabular}{cp{2.7in}}
        \toprule
        \textbf{Type} & \textbf{Description}\\ \midrule
        I & Identical sans whitespace and comments\\
        II & Identical AST but uses different variable names, types or function calls\\
        III & Similar AST but uses different expressions/statements. For example, a) using \mintinline{Java}{while} in place of \mintinline{Java}{for} loops or b) using \mintinline{Java}{if else if} in place of \mintinline{Java}{switch} statements.\\
        IV & Different syntax but behaviorally same. For example, an iterative stack approach or a recursive approach can be used for breadth first search of a graph.\\
        \bottomrule
    \end{tabular}
    \label{tab:types}
\end{table}

We propose \longapproach{} (\approach{}), a semantic approach to code similarity that is predicated on the availability of large repositories of redundant code~\cite{barr2014plastic}. Instead of mapping API translations using predefined rules~\cite{java2Csharp, bowles1983multi}, or using embedded API translations~\cite{nguyen2017exploring, beit2002source}, \approach{} uses IO examples to cluster code based on its behavior. Further, it relaxes the bounds of the datatypes across programming languages, which helps dynamic typed code snippets (e.g., Python) to be clustered alongside static typed code snippets (e.g., Java). 

\begin{figure}[tb]
    \centering
    \includegraphics[width=\linewidth]{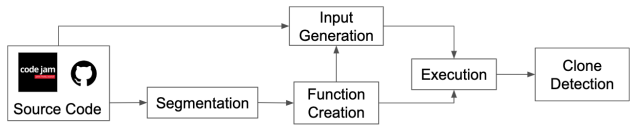}
    \caption{High level workflow for \approach{}.}
    \label{fig:slacc}
\end{figure}

In \approach{}, we build on the ideas pioneered by EQMiner~\cite{jiang2009automatic} for using segmentation and random testing for clone detection.
\approach{} starts by identifying  snippets from a large code base and involves a multi-step process depicted in \figref{slacc}, which starts with a)~\textit{Segmentation} of the code base into smaller fragments of code called {snippets}, b)~\textit{Function} \textit{creation} from the snippets, c)~\textit{Input} \textit{generation} for the functions, d) \textit{Execution} of the functions, and e)~\textit{Clone} \textit{detection} based on clustering functions arguments and execution results. 


\subsection{Segmentation}
\label{sect:snip}

In the first stage, code from all the source files in a project is broken into smaller code fragments called \emph{snippets}. Consecutive statement blocks of threshold \mintinline{Java}{MIN_STMT} or more are grouped into a snippet. A statement block can be
\begin{enumerate}
    \item \textbf{Declaration} Statement. \textit{e.g.,}  \mintinline{Java}{int x;}
    \item \textbf{Assignment} Statement \textit{e.g.,}  \mintinline{Java}{x = 5;} 
    \item \textbf{Block} Statement  \textit{e.g.,} \mintinline{Java}{static {x = 10;}}
    \item \textbf{Loop} statements. \textit{e.g.,} \mintinline{Java}{for}, \mintinline{Java}{while}, \mintinline{Java}{do-while}
    \item \textbf{Conditional} statements. \textit{e.g.,} \mintinline{Java}{if}, \mintinline{Java}{if-else-if}, \mintinline{Java}{switch},
    \item \textbf{Try} Statement. \textit{e.g.,} \mintinline{Java}{try}, \mintinline{Java}{try-catch}
\end{enumerate}

\noindent \aref{segment} illustrates the segmentation phase. For an AST $A_F$ of a function, the algorithm performs a pre-order traversal of all the nodes in the AST (\textit{line 5}) and then uses a sliding window to extract segments of size greater than a minimum segment size \mintinline{Java}{MIN_STMT} (\textit{lines 12-13}). Further, for statements like Block, Loop, Conditional and Try which have statements in its nested scope, the algorithm is called recursively on them (lines 14-15).

\subsection{Function Creation}
\label{sect:type}

Next, snippets are converted into executable functions. This section describes how arguments, return variables, and types are inferred. 



\begin{algorithm}[tb]
\caption{Segmentation}\label{alg:segment}
\begin{algorithmic}[1]
\State \textbf{Input:} $A_F$ - AST Node
\State \textbf{Output:} $\mathbb{S}$ - List of Segment
\Procedure{Segment}{$A_F$}
    \State $\mathbb{S} \gets \phi$
    \State stmts $\gets \, PreorderTraverse(A_F)$
    \ForAll {$i \, \in \, $range$(0, \, $\textbf{len}$($stmts$) - 1)$} 
        \State $S_i \gets \{\}$ 
        \State $stmt_i \gets \,$stmts$[i]$ 
         \ForAll {$j \, \in \, $range$(i, \, $\textbf{len}$($stmts$))$}
            \State $stmt_j \gets $stmts$[j]$ 
            \State $S_i.$append$(stmt_j)$ 
            \If {\textbf{len}$(S_i) \geq MIN\_STMTS $}
                \State $\mathbb{S} \gets \mathbb{S} \cup S_i$
            \EndIf
            \If {$stmt_j.$\textbf{hasChildren}$()$}
                \State $\mathbb{S} \, \gets \, \mathbb{S} \, \cup \,  $\textbf{SEGMENT}$(stmt_j)$
            \EndIf
         \EndFor
    \EndFor
    \State \Return $\mathbb{S}$
\EndProcedure
\end{algorithmic}
\end{algorithm}
 
\paragraph{Inferring arguments and return variables}
We adapt a dataflow analysis similar to that used by Su \etal ~\cite{su2016identifying}. 
For each method, potential return variables are identified as variables that are defined or modified within the scope of the snippet. 
If the last definition of a variable is a constant value, that variable is removed from the set of potential return variables.  
Arguments are variables that are 1) used but not defined within the scope of the snippet, and 2) not declared as public static variables for the class. 
For each potential return variable in a snippet, a function is created. 

\paragraph{Inferring types}
In the case of static typed languages, argument types and return values can be inferred via static code analysis. For dynamic typed languages, the parameters can take multiple types of input arguments. This  increases the possible values of the arguments generated (see \sect{args}) to identify its behavior. In many cases, the possible types for the arguments can be inferred by parsing the code and looking for constant variables~\cite{cousot1977abstract} in its context. This technique has been used in inferring types in other dynamic languages like JavaScript~\cite{jensen2009type}. For example, in the following Python function, the type of \mintinline{Python}{n} can be  assumed to be an integer since it is compared against an integer. 

\begin{minted}[xleftmargin=2em,linenos,firstnumber=1,fontsize=\small]{Python}
def fib(n):
    if n <= 1: return n
    return fib(n-1) + fib(n-2)
\end{minted}

\noindent In cases where the types of the parameters could not be inferred at compile time, such as:
\begin{minted}[xleftmargin=2em,linenos,firstnumber=1,fontsize=\small]{Python}
def main(a):
    print a
\end{minted}
\noindent a generic type is assigned (i.e., for \mintinline{Python}{a}) allowing the argument to assume any of the primitive types used in argument generation (\sect{args}).

\paragraph{Converting object return types into functions}
If a snippet returns an object, the object is simplified into multiple functions returning each of its non-private members independently. For example, in \figref{objectRets}, \mintinline{Java}{func_s} has a return type of \mintinline{Java}{Shape}. Shape has two members, \mintinline{Java}{length} and \mintinline{Java}{width}. Hence, \mintinline{Java}{func_s} is broken down into two functions, \mintinline{Java}{func_l} and \mintinline{Java}{func_w}, which return the \mintinline{Java}{length} and \mintinline{Java}{width} of the shape object independently. Note that a third function for \mintinline{Java}{height} is not created since it is a private member.

\begin{figure}[tb]
    \begin{minted}[xleftmargin=2em,linenos,firstnumber=1,fontsize=\small]{Java}
class Shape {
    public int length;
    int width;
    private int height;
    public Shape(int l, int w, int h) {
        length=l; width=w; height=h;
    }
}
public Shape func_s (int l, int w, int x) {
    return new Shape(l + x, w * 2, x);
}
public int func_l (int l, int w, int x) {
    return func_s(l, w, x).length;
}
public int func_w (int l, int w, int x) {
    return func_s(l, w, x).width;
}
    \end{minted}
    \caption{An example depicting conversion of a function with object as a return type to multiple functions with non-primitive members of the object's class.}
    \label{fig:objectRets}
\end{figure}

\paragraph{Permuting argument order}
For each of the snippets, we generate different permutations based on the input of arguments since order matters for capturing function behavior. 
Consider the two functions in \figref{inputReorder}; the first function divides  \mintinline{Java}{a} with \mintinline{Java}{b} using the division (\mintinline{Java}{/}) operator while the second divides \mintinline{Java}{dividend} with \mintinline{Java}{divisor} using the subtract (\mintinline{Java}{-}) operator recursively. For the inputs (5, 2) the two functions would produce the values 2 and 0 respectively. But if the arguments for the second function was reversed, it would produce the same output 2. Thus, for every function, we create duplicates in different permutations of the arguments, \mintinline{Java}{ARGS}, resulting in $\lvert$\mintinline{Java}{ARGS}$\rvert !$ different functions. To limit the creation of this exploding space, we set an upper limit on the number of arguments per function that is included in the analysis (\mintinline{Java}{ARGS_MAX}). 

\subsection{Input Generation}
\label{sect:args}

A set of inputs are required to execute the created functions. Following this, clustering is performed.

\paragraph{Input creation} 
Inputs are generated based on argument type and using a custom input generator inspired by grey-box testing~\cite{khan2012comparative} and multi-modal distribution~\cite{jiang2009automatic}. First, the source code is parsed and constants of each type are identified. Next, a multi-modal distribution is declared for each of the types with peaks at the constants. Finally, values for each type are sampled from this multi-modal distribution. Our experiments create 256 inputs per function, as justified in \sect{inputVariance}.

\paragraph{Memoization} For every function with the same argument types, a common set of inputs have to be used to compare them. This is ensured using a database and the input generator. The generator is used to create sample inputs for the given argument types and stored in the database. For subsequent functions with the same signature for the arguments, the stored input values are reused.

\begin{figure}[tb]
    \begin{minted}[xleftmargin=2em,linenos,firstnumber=1,fontsize=\small]{Java}
public int divide_simple (int a, int b) {
    if (b == 0) return 0
    return a / b;
}
public int divide_complex (int divisor, int dividend) {
    // Same as dividend/divisor
    if (b == 0) return 0
    int quotient = 0;
    while (dividend >= divisor) {
        dividend = dividend - divisor;
        quotient++;
    }
    return quotient;
}
    \end{minted}
    \caption{An example illustrating the need for reordering arguments. The two functions perform integer division but do not return the same return value for the same set of inputs due to the order of arguments in the function definition.}
    \label{fig:inputReorder}
\end{figure}


\paragraph{Supported argument types} \approach{} currently supports four types of arguments. 
\begin{enumerate}
    \item \textbf{Primitive}. The multi-modal distribution for the argument type is sampled to generate the inputs. This includes integers (and longs, shorts), floats (and double), characters, booleans, and strings.
    \item \textbf{Objects}. Objects are recursively expanded to their constructor with primitive types; inputs are generated for the types. 
    \item \textbf{Arrays}. A random array size is generated using the input generator for integers\footnote{If a negative integer is sampled, the distribution is re-sampled.}. For each element in the array, a value is generated based on the array type (Primitive or Object).
    \item \textbf{Files}: Files are stored as a shared resource pool of strings in the database. If a seed file(s) is provided, it is randomly mutated and stored as a string in the database. In the absence of a seed, constants from the multi-modal distribution are sampled and stored as strings. For an argument with a \mintinline{Java}{File} type (or its extensions), a temporary (deleted on termination) file object is created using the stored strings.
\end{enumerate}

\paragraph{Type size restrictions} Comparing code snippets requires compatible sizes of types across programming languages. For example, Java has 4 integer datatypes \mintinline{Java}{byte}, \mintinline{Java}{short}, \mintinline{Java}{int} and \mintinline{Java}{long} which occupy sizes of 1, 2, 4 and 8 bytes, respectively. On the other hand, Python has two integer datatypes: \mintinline{Python}{int} which is equivalent to the \mintinline{Java}{long} datatype in Java and \mintinline{Python}{long} which has an unlimited length. Thus, we make a restriction when generating inputs for functions across different languages: inputs are generated from the smaller bound of the two programming languages. For example, in the case of Java and Python function that has an \mintinline{Java}{int}, inputs are generated within the bounds of Java.


\subsection{Execution}
\label{sect:exec}
In the next stage, the created functions are executed over the generated input sets and the subsequent return values are stored. Each function is assigned an execution time limit of $T_L$ seconds, after which a Timeout Exception is raised. This occurs most frequently when there is an infinite loop, such as \mintinline{Java}{while(true)} when the loop invariant is an argument. Each execution of the function is run on an independent thread. Subsequently, the return value, runtime and exception for the executed function over the input set is stored.

\subsection{Clone Detection}
\label{sect:cluster}

The last stage of \approach{} is identifying the clones, where the executed functions are clustered on their inputs and outputs. 
\approach{} uses a \textit{representative based partitioning strategy}~\cite{roy2009comparison, su2016identifying} to cluster the executed functions.

\paragraph{Similarity Measure}
In this work, a pair of functions have the highest semantically similarity if for any given input, the functions return the same output. The similarity measure between two functions is computed as the number of inputs for which the methods return the same output value divided by the number of inputs, same as the Jaccard index. This creates a similarity value between two functions with a range of [$0.0,1.0$] with 1.0 being the highest.


Consider the functions from \sect{motivation}, \mintinline{Java}{interleave},\linebreak \mintinline{Python}{fancy_interleave}, and
\mintinline{Python}{valid_interleave}. For values \mintinline{Java}{a = [2,3]} and \mintinline{Java}{b = [4]}, we see that 
\mintinline{Java}{interleave(a,b) = [2,4,3]}, \linebreak
\mintinline{Java}{fancy_interleave(a,b) = [2,4]} 
and \mintinline{Java}{valid_interleave(a,b)}\linebreak \mintinline{Java}{ = [2,4,3]}. Functions 
\mintinline{Java}{interleave} and \mintinline{Java}{valid_interleave} are similar since they have the same output for the same input but \mintinline{Java}{interleave} and \mintinline{Java}{fancy_interleave} are not similar. In contrast, for \mintinline{Java}{a = [2,3]} and \mintinline{Java}{b = [4,5]}, all three functions would return the same output \mintinline{Java}{[2,4,3,5]}.
Based on these two inputs, \mintinline{Java}{interleave} and \mintinline{Python}{fancy_interleave} have a similarity of $0.5$,  \mintinline{Java}{interleave} and \mintinline{Python}{valid_interleave} have a similarity of $1.0$, and \mintinline{Python}{fancy_interleave} and \mintinline{Python}{valid_interleave}  have a similarity of $0.5$. 
This process is repeated for many such inputs \mintinline{Java}{a} and \mintinline{Java}{b} to compute similarity scores between each pair of functions. 


 Functions are only compared if they have the same number of arguments and  cast-able argument types. For example, consider the four functions \mintinline{Java}{f1(int a, String b)},  \mintinline{Java}{f2(long a, File b)}, \mintinline{Java}{f3(File a, String b)} and \mintinline{Java}{f4(String a)}. Functions \mintinline{Java}{f1} and \mintinline{Java}{f2} can be compared since \mintinline{Java}{int} can be cast to a \mintinline{Java}{long} value. But they cannot be compared to \mintinline{Java}{f3} since primitive types cannot be cast to \mintinline{Java}{File}. Similarly, \mintinline{Java}{f1}, \mintinline{Java}{f2} and \mintinline{Java}{f3} cannot be compared \mintinline{Java}{f4} due to the difference in number of arguments.
 
 

\begin{algorithm}[tb]
\caption{Clustering}\label{alg:cluster}
\begin{algorithmic}[1]
\State \textbf{Input:} $\mathbb{F}$ - List of Functions with Input and Output
\State \textbf{Output:} $\mathbb{C}$ - List of clusters
\Procedure{Cluster}{$\mathbb{F}$}
    \State $\mathbb{C} \gets \phi$
    \ForAll {$F \in \mathbb{F}$} 
         \ForAll {$C \in \mathbb{C} $}
            \State $O \gets \textbf{GetRepresentive}(C)$
            \If {$\textbf{Similarity}(O, F) \geq SIM\_T $}
                \State $C \gets C \cup F$
                \State \textbf{break}
            \EndIf
         \EndFor
         \If {$\forall C \in \mathbb{C}, F \notin C$}
            \State $C_{|C|+1} \gets {F}$
            \State $\textbf{SetRepresentative}(C_{|C|+1}, F)$
            \State $\mathbb{C} \gets \mathbb{C} \cup C_{|C|+1}$ 
         \EndIf
    \EndFor
    \State \Return $\mathbb{C}$
\EndProcedure
\end{algorithmic}
\end{algorithm}

\paragraph{Clustering}

A function is compared to a cluster by measuring its similarity with the first function added to the cluster (called \emph{representative}). The clustering algorithm is briefly described in \aref{cluster}. An empty set of clusters is first initialized (\textit{line 4}). Each function (\textit{line 5}) is compared against each cluster (\textit{line 6}). If the similarity between the  \emph{representative} (\textit{line 7}) and the function is greater than a predefined similarity threshold, \mintinline{Java}{SIM_T} (\textit{line 8}), the function is added to the cluster (\textit{line 9}). If the function does not belong in any cluster (\textit{line 11}), a singleton cluster is created for the function (\textit{line 12}) and the function is set as the cluster's \emph{representative} (\textit{line 13}). The singleton cluster is added to the set of clusters (\textit{line 14})


\section{Evaluation}
\label{sect:eval}

Our goal is to evaluate the effectiveness of \approach{}. 
There is a three-phase evaluation, first to compare \approach{} to a comparable technique in a single, static typed language. Next, we apply \approach{} to a single, dynamic typed language (Python) and then to a multi-language context; in both cases \approach{} is compared to type-III clones.

\subsection{Research Questions}

\approach{} is benchmarked against \baseline{}~\cite{su2016identifying} with respect to coverage and precision of code-clone detection. This leads us to our first research question:

\begin{resQues}[1]
How effective is \approach{} on semantic clone detection in \added{static} typed languages?
\end{resQues}

\noindent Prior research has already shown that semantic clones can be found in static typed languages~\cite{jiang2009automatic, elva2012jsctracker, su2016identifying} like C and Java. In our literature search, we failed to find techniques that identified semantic code clones in dynamic typed languages.  Therefore, we use an AST based comparison approach as an alternative baseline to benchmark \approach{}. This leads us to the next research question: 

\vspace{1.0in}

\begin{resQues}[2]
How effective is \approach{} on semantic clone detection in dynamic typed languages?
\end{resQues}

\noindent Prior work identified code clones between languages by mapping APIs between similar languages (e.g., Java and C\#) using predefined rules~\cite{java2Csharp} or using an embedded API translations~\cite{nguyen2017exploring, beit2002source}. As a result, these code clones are syntactic rather than semantic. Therefore:

\begin{resQues}[3]
How effective is \approach{} at cross-language semantic clone detection?
\end{resQues}


\subsection{Data}
We validate this study on four problems from Google Code Jam (GCJ) repository and their valid submissions in Java and Python. GCJ is an annual online coding competition hosted by Google where participants solve the
programming problems provided and
submit their solutions for Google to test. The
submissions that pass Google's tests are considered valid and are published online. 
We use the first problem from the fifth round of GCJ from 2011 to 2014\footnote{Early rounds have many submissions to create a reasonably scoped experiment. Thus, we chose submissions from the quarterfinals in round five.}. The details about the problem and submissions are in \tref{data}. Overall in this study, we consider 247 projects; 170 from Java and 77 from Python. 
The 170 Java GCJ submissions contain 885 methods and generated 19,188 Java functions.
 The 77 Python submission contains 301 methods and generated 17,215 Python functions. 
 

\begin{table}[tb]
    \centering
    \caption{Projects used in this study with the number of valid submissions in both Java and Python.}
    \begin{tabular}{ccccc}
        \toprule
         \textbf{Year} & \textbf{Problem} & \textbf{ID} & \textbf{Java} &  \textbf{Python} \\ \midrule
         2011 & Irregular Cake & Y11R5P1 & 48 & 16 \\
         2012 & Perfect Game & Y12R5P1 & 47 & 24 \\
         2013 & Cheaters & Y13R5P1 & 29 & 19 \\
         2014 & Magical Tour & Y14R5P1 & 46 & 18 \\ \midrule
         & \textbf{Total} & & 170 & 77 \\ \bottomrule
    \end{tabular}

    \label{tab:data}
\end{table}

The code, projects and execution scripts for the project can be found in our {GitHub Repository}~\cite{slaccTool}.

\subsection{Experimental Setup}
\label{sect:setup}

The experiments were run on a 16 node cluster with each node having a 4-core AMD opteron processor and 32GB DDR3 1333 ECC DRAM. 
Our experiments have four hyper-parameters
\begin{itemize}
    \item Minimum size of snippet (\mintinline{Java}{MIN_STMT} - \sect{snip}): We set this to 2 to \added{capture snippets with interesting behavior}. 
    \item Maximum number of arguments (\mintinline{Java}{ARG_MAX} - \sect{type}): This value is set to 5. Hence if a snippet has more than 5 arguments, it is omitted from the experiments.
    \item Number of executions (\sect{exec}): We execute each snippet with 256 generated inputs (\sect{args}); see \sect{inputVariance} for details on this choice.
    \item Similarity Threshold (\mintinline{Java}{SIM_T} - \sect{cluster}): We set this to 1.0 for our experiment. This implies that two functions are only considered to be clones if for \tbi{all} inputs they generate the same outputs.
\end{itemize}

Sensitivity to the number of executions and \mintinline{Java}{ARG_MAX} is explored and discussed in Sections ~\ref{sect:inputVariance} and ~\ref{section:argmax} respectively.

\subsection{Metrics}

Our study uses three metrics primarily to address the research questions we pose.
\begin{itemize}
    \item \textbf{Number of Clusters}: A cluster is a collection of functions with a common property (i.e., type I-IV similarity). This metric is the number of clusters generated by a clone detection algorithm. This is represented as |Clusters|, \# Clusters or \#C.
    \item \textbf{Number of Clones}: A function that belongs to a cluster is called a clone. This metric is the total number of functions in all the clusters generated by a clone detection algorithm. This is represented as |Clones|, \# Clones or \#M.
    \item \textbf{Number of False Positives}: A false positive is a cluster which contains one or more functions which does not adhere to the similarity measure of the cluster. This is represented as |False Positive|, \# False Positives or \#FP.
\end{itemize}

\subsection{Baselines}
\label{sect:baseline}
To answer RQ1, RQ2, and RQ3, we use baseline techniques to illustrate the capabilities of \approach{}.

\subsubsection{RQ1: \baseline{}}
\label{sect:Hitoshi}
As a baseline, we use the closest technique to ours, 
\baseline{}~\cite{su2016identifying}. This tool identifies functional clones for Java Virtual Machine (JVM) based languages such as Java and Scala. 
It uses in-vivo clone detection and 
inserts instrumentation code in the form of control instructions~\cite{javaByteCode} in the application's bytecode to record input and output values at runtime. 
Inputs and outputs are observed using the existing workloads, which allows it to observe behavior and identify clones in code for which input generators cannot generate inputs. 
The methods with similar values of inputs and outputs during executions are identified as functional clones.  \baseline{} considers every method in a project as a potential functional clone of every other method 
and returns pairs of clones. For comparison against \approach{},  we group the pairs into clusters as follows: two pairs of clones are grouped into a cluster if both the pairs have a common function between them (i.e., for pairs (A,B) and (B,C), a clone cluster is created with (A,B,C)).

Like the similarity threshold \mintinline{Java}{SIM_T} in \approach{}, \baseline{} has a similar parameter that provides a lower bound on how similar two methods must be to be considered a functional clone.  As with \approach{}, \baseline{} also has a parameter for an upper bound on the number of IO profiles considered for each method.  

We used an existing and public implementation of \baseline{}.\footnote{\href{https://github.com/Programming-Systems-Lab/ioclones}{github.com/Programming-Systems-Lab/ioclones}; 

\noindent Commit hash: \href{https://github.com/Programming-Systems-Lab/ioclones/tree/aa5b5b3ed7fe311564ba1508b1b22fb47ccc2979}{aa5b5b3}; Dated: 05/06/2018}
The workload used to benchmark \baseline{} with GCJ are the sample test input files. GCJ provides only two sample input files for a validating a submission. However, in \approach{} each method was executed 256 times. To create a balanced benchmark, we randomly fuzzed the test input files 32 times before sending it to \baseline{}. Note that we tried fuzzing the files 256 times but the clone-detection phase of \baseline{} crashed for large numbers of inputs. 



\subsubsection{RQ2: Automated AST Comparison}
\label{sect:AST}

To the best of our knowledge we could not find a prior work to detect semantic code clones in dynamic languages. Hence we benchmarked \approach{} for dynamic and cross-language clones by matching the Abstract Syntax Trees (ASTs) as a proxy for similarity. This technique has been adopted by many graph-based (an example of type-III clone) code clone detection techniques in C~\cite{baxter1998clone, yang1991identifying,jiang2007deckard} and Java~\cite{jiang2007deckard, koschke2006clone}.

Like \approach{}, the first phase of the AST comparison segments the code into snippets. Next an AST is generated for the snippets. We use the JavaParser~\cite{javaParser} tool and Python AST~\cite{pythonAST} module to construct the ASTs in the respective languages. We measure similarity by matching the ASTs. For clones in the same programming language (RQ1, RQ2), we match the ASTs and consider them to be type-III clones if the ASTs are equivalent or have a difference of at most one node. 

\subsubsection{RQ3: Manual Cross-language AST Comparison}


The \linebreak automated AST comparison approach cannot be adopted for cross-language clones (RQ3) due to the difference in format of the ASTs for both the languages. 
In this case, conservatively, we sampled cross-language snippets with extremely similar outputs and manually verified the ASTs for similarity. 
To do this, we randomly sample $1$ million pairs of a Java function and a Python function. If the input and output types are compatible, and the outputs are the same for the same inputs or off by a \emph{consistent} value, then we manually evaluate the ASTs for similarity. \emph{Consistency} is determined based on the output type. Values of primitive types are consistent if they have a constant difference (for Boolean or Numeric values), constant ratio (for Boolean or Numeric values) or constant Levenshtein distance~\cite{wiki:levenshtein} (for Strings) between the outputs. Objects are consistent if each member of the object is consistent. Finally, two arrays are consistent, if all the corresponding members of the array are consistent.

For example, given two methods, \mintinline{Java}{int A(int x)} and \mintinline{Python}{def B(y)}, if \mintinline{Java}{A(1) = 1}, \mintinline{Python}{B(1) = 9}, \mintinline{Java}{A(2) = 2}, and \mintinline{Python}{B(2) = 18}, then \mintinline{Java}{A()} and \mintinline{Python}{B()} are similar since their outputs have a constant ratio ($9$).  Of the 616 similar pairs, all had identical ASTs or had a difference of at most one node, making them type-III clones. 

\subsection{Precision Analysis}
\label{sect:fp}
\approach{} and \baseline{} are both clustered using IO relationships of the functions. However, given a different set of inputs, some functions in a cluster might produce a different set of outputs such that they are not clones; such clusters are marked as \textit{false positives} and considered invalid. We identify false positives at the cluster-level in keeping with prior work~\cite{jiang2009automatic}.

To detect false positives, \approach{} clusters are re-executed on a new set of 256 inputs generated using random fuzzing~\cite{jiang2009automatic} based on a triangular distribution, and clustered. If any method in a cluster is not grouped into the same cluster using the new input set, the whole cluster is marked as a false positive. We observe that the number of clusters and false positives is relatively stable above 64 inputs (\sect{inputVariance}). 

To detect false positives in \baseline{}, we randomly fuzz the test input files 32 times (\sect{baseline}) to generate a new test file that is 32x the size of the original, and then re-execute \baseline{}. Clone pairs are clustered and false-positives are detected when a new cluster does not match an original cluster, as done for \approach{}. 

False positives in clusters generated by AST comparisons are identified in a similar manner to \approach{}. ASTs in the clusters are first converted to functions (as described in \sect{type}). The functions are re-executed on 256 inputs like \approach{} clusters and checked for false positives. Any cluster that contains a different method after execution is marked as a false positive.

\section{Results}

The results show that \approach{} identifies more method level clones compared to prior work and with higher precision (RQ1), successfully identifies clones in dynamic typed languages (RQ2), and successfully detects clones between Java and Python (RQ3).

\subsection{RQ1: Static Typed Languages}

\newtcolorbox{codebox}[2][]{%
  enhanced,colback=white,colframe=black,coltitle=black,
  sharp corners,boxrule=0.4pt,
  fonttitle=\itshape,
  left=2pt,right=2pt,top=2pt,bottom=2pt,
  attach boxed title to top left={yshift=-0.3\baselineskip-0.4pt,xshift=2mm},
  boxed title style={tile,size=minimal,left=0.5mm,right=0.5mm,
    colback=white,before upper=\strut},
  title=#2,#1
}

\begin{figure}[tb]
    \begin{codebox}{\approach{}\textsubscript{stmt}}
        \begin{minted}[xleftmargin=2em,linenos,firstnumber=1,fontsize=\small]{Java}
import Y14R5P1.stolis.MMT3 // Parent Class MMT3
public static int func_a(BufferedReader br){
  // Snipped from Y14R5P1.stolis.MMT3.main() 
  if (!MMT3.in.hasMoreTokens())
    MMT3.in = new StringTokenizer(br.readLine());
  int a = Integer.parseInt(MMT3.in.nextToken());
  return a;
}
        \end{minted}
        \begin{codebox}{\approach{}\textsubscript{method}}
            \begin{minted}[xleftmargin=2em,linenos,firstnumber=1,fontsize=\small]{Java}
import Y12R5P1.xiaowuc.A // Parent Class A
public static int func_b(Scanner in) {
  // Y12R5P1.xiaowuc.A.next()
  while (A.tok == null || !A.tok.hasMoreTokens()) {
    A.tok = new StringTokenizer(in.readLine());
  }
  return Integer.parseInt(A.tok.nextToken());
}
            \end{minted}
            \begin{codebox}{\baseline{}}
                \begin{minted}[xleftmargin=2em,linenos,firstnumber=1,fontsize=\small]{Java}
public static int func_c(StreamTokenizer in) {
  // Y11R5P1.burdakovd.A.nextInt()
  in.nextToken();
  return (int) in.nval;
}
                \end{minted}    
                \begin{minted}[xleftmargin=2em,linenos,firstnumber=1,fontsize=\small]{Java}
public static int func_d(StreamTokenizer in) {
  // Y11R5P1.Sammarize.Main.next()
  in.nextToken();
  return Integer.parseInt(in.nval);
}
                \end{minted}
            \end{codebox}
        \begin{minted}[xleftmargin=2em,linenos,firstnumber=1,fontsize=\small]{Java}
import Y14R5P1.eatMore.A // Parent Class A
public static int func_e(Scanner in) {
  // Y14R5P1.eatMore.A.next()
  A.in = in;
  return Integer.parseInt(A.nextToken());
}
        \end{minted}
        \end{codebox}
        \begin{minted}[xleftmargin=2em,linenos,firstnumber=1,fontsize=\small]{Java}
public static int func_f(Scanner sc) {
  // Snipped from Y11R5P1.dooglius.A.go() 
  int next = sc.nextInt();
  return next;
}
        \end{minted}
    \end{codebox}
    \caption{Semantic clusters detected by HitoshiIO, \approach{} on method level (\approach{}\textsubscript{method}) and \approach{} on statement level (\approach{}\textsubscript{stmt}). The cluster contains functions that take an object that reads a file and returns the next Integer token. }
    \label{fig:rq1Clones}
\end{figure}




The 885 Java methods generated 19,188 Java functions for analysis. \approach{} was able to support 691 of the 885 Java methods. From the 691 whole methods, 18,497 functions are derived into partial method snippets. Of the total generated functions, 4,180 (22\%) are clones resulting in 632 clusters. 
These 4,180 clones derive from 4,038 partial-method snippets and 142 whole methods. We call them \emph{statement level} clones and \emph{method level} clones, respectively. 

\subsubsection{Method level clones}
\label{sect:methodClonesRQ1}

\begin{table}[tb]
    \centering
      \caption{Number of whole method clones identified by HitoshiIO($H$), \approach{}($S$) and both the approaches, after accounting for false positives.}
    \begin{tabular}{cccc}
        \toprule
        \textbf{Problem} & \textbf{\baseline{}($\lvert$H$\rvert$)}& \textbf{\approach{}($\lvert$S$\rvert$)} & \textbf{$\lvert$H$\cap$S$\rvert$}\\
        \midrule
        Irregular Cake & 3 & 44 & 3\\
        Perfect Game  & 4 & 35 & 4\\
        Cheaters  & 4 & 21 & 4\\
        Magical Tour  & 9 & 35 & 9\\ \midrule
        \textbf{Total}& 20 & 135 & 20\\ \bottomrule
    \end{tabular}
    \label{tab:wholeMethodCoverage}
\end{table}

We benchmark \approach{} against \baseline{} by comparing clones detected by \approach{} at a method level granularity.
We provide all 885 Java methods to \baseline{}, which groups 43 of the methods into 13 clusters.  False positives were identified for 9 of the 13 clusters (precision=30.7\%).\footnote{False positive rates in the original \baseline{} paper~\cite{su2016identifying} are computed at the pair-level rather than cluster level and used student opinions rather than code behavior, which may account for the relatively low precision reported here.} 
The remaining valid clusters from \baseline{} contain 20 methods. 
From the 691 Java methods, \approach{} detected 142 methods, grouped into 15 clusters. False positives were identified for 2 of the 15 clusters (precision = 86.7\%). The remaining valid clusters for \approach{} contain 135 methods. 

~\tref{wholeMethodCoverage} shows the numbers of valid clusters for each approach, as well as their intersection. All valid clusters from \baseline{} are contained within the valid clusters for \approach{}, ($H \equiv H \cap S$), demonstrating that among the valid clones, \approach{} subsumes \baseline{} for this experiment. However, the low precision for \baseline{} may be due to the use of limited inputs or the execution context, so further investigation is needed for generalization of this result. 


An example of a cluster that contains methods from both \approach{} and \baseline{} is shown in \figref{rq1Clones}. The cluster contains functions that take an object that reads a file and returns the next \mintinline{Java}{Integer} token. Functions \mintinline{Java}{func_c} and \mintinline{Java}{func_d} are clones detected by \baseline{}. Within the same cluster, \approach{}\textsubscript{method} additionally identifies two more method level clones that were not detected by \baseline{}: \mintinline{Java}{func_b} and \mintinline{Java}{func_e}. 

\subsubsection{Statement level clones}
\label{sect:statClonesRQ1}

Additionally, \approach{} identifies 624 clusters with 4,038 statement level code clones. Of these, 48 clusters are false positives (precision=92.3\%). The large number of code clones is intuitive because each method can contain multiple modular functionalities. That said, it should be noted that the higher precision for statement level clusters would lead us to believe that detecting clones for succinct behavior is more accurate. 

Statement level clones can be clustered with whole method clones. For example, in \figref{rq1Clones}, \approach{}\textsubscript{stmt} represents a \approach{} cluster based on partial methods: \mintinline{Java}{func_a} and \mintinline{Java}{func_f} are functions segmented from the \mintinline{Java}{main} method in class \mintinline{Java}{Y14R5P1.stolis.MMT3} and the \mintinline{Java}{go} method in  \mintinline{Java}{Y11R5P1.dooglius.A}, respectively. 

\begin{tcolorbox}
\textbf{RQ1:} \emph{Method level clones:} \approach{} identifies more method level clones compared to \baseline{} at higher precision. 
\emph{Statement level clones:} Segmentation of code increases the precision of \approach{} and yields a higher number of semantic clones. 
\end{tcolorbox}

\begin{table}[tb]
    \centering
     \caption{
        \# of Java, Python and Cross language clusters detected by \approach{} compared against AST (Type-III) clusters.
    }
    \begin{tabular}{lcccccc}
        \toprule
        & \multicolumn{2}{c}{\textbf{Java}} & \multicolumn{2}{c}{\textbf{Python}} & \multicolumn{2}{c}{\textbf{Java + Python}}\\ \cmidrule{2-3}\cmidrule{4-5}\cmidrule{6-7}
        & \approach{} & AST & \approach{} & AST & \approach{} & AST \\ \midrule
        \textbf{\# Clusters} & 632 & 6122 & 482 & 3971 & 34 & 616 \\
        \textbf{\# Valid} & 584 & 226 & 421 & 181 & 32 & 25\\
        \textbf{Precision} & 92.4 & 3.7 & 87.3 & 4.6 & 94.1 & 4.1 \\ \bottomrule
    \end{tabular}
   
    \label{tab:type3Validity}
\end{table}

\subsection{RQ2: Dynamic Typed Languages}
\label{sect:ra1}



\approach{} identified that 3,135 (18.2\%) of the 17,215 extracted Python functions had clones which resulted in 482 clone clusters.  
Of these 482 clusters, 421 are valid, resulting in precision of 87.3\%. 
As a baseline, using the same Python functions, we systematically looked for type-III clones. 
There exists 3,971 clusters, of which 181 are valid (4.6\% precision); these results are shown in the \emph{Python} column of \tref{type3Validity}, where \emph{AST} shows the type-III clones. 
For sake of comparison, the experiment was repeated for Java clones; a similar differential between \approach{} and AST precision was observed (92.4\% vs. 3.7\%).

 When these clusters are validated, 61 of the 482 \approach{} clusters (12.8\%) were deemed to be false positive. This is more than the percentage of false positives in Java ($7.3\%$), but we suspect that by executing the functions over a larger set generated arguments, the subsequent clustering could yield more robust results.

An example of Python clones identified by \approach{} can be seen in \figref{pySimions}. Both the functions in this example compute the sum of an array. \mintinline{Python}{func_db8e} uses a loop that maintains the running sum where each index in the array contains the array sum until that index. The last index of the array would contain the array sum and is eventually returned. In contrast, \mintinline{Python}{func_43df} uses the \mintinline{Python}{sum} library function to perform the same task. 

\begin{figure}[tb]
    \begin{minted}[xleftmargin=2em,linenos,firstnumber=1,fontsize=\small]{Python}
def func_db8e(a):
    n = len(a)
    sum0 = [0] * (n + 1)
    for i in xrange(n):
        sum0[i + 1] = sum0[i] + a[i]
    allv = sum0[-1]
    return allv
\end{minted}    
\begin{minted}[xleftmargin=2em,linenos,firstnumber=1,fontsize=\small]{Python}
def func_43df(items):
    _sum = sum(items)
    j = len(items) - 1
    return _sum
\end{minted}    
    \caption{Semantic cluster of Python functions detected by \approach{}. The cluster contains functions that returns the sum of an input array.}
    \label{fig:pySimions}
\end{figure}


\begin{tcolorbox}
\textbf{RQ2:} \approach{} can successfully identify code clones for dynamic typed languages with high precision (87.3\%). 
\end{tcolorbox}


\begin{figure}[tb]
\begin{minted}[xleftmargin=2em,linenos,firstnumber=1,fontsize=\small]{Java}
static long func_3b0e (Long[] x2) {
    Long res = null;
    Long[] arr = x2;
    int len = arr.length;
    for (int i = 0; i < len; ++i) {
        long xx = arr[i];
        if (xx >= res)
            continue;
        res = xx;
    }
    return res;
}
\end{minted}    

\begin{minted}[xleftmargin=2em,linenos,firstnumber=1,fontsize=\small]{Python}
def func_6437 (y):
    ymin = min (y)
    count = 0
    return ymin
\end{minted}    
    \caption{
    Semantic cluster of a Java function and a Python function detected by \approach{}. The cluster contains functions that returns the minimum value in an input integer array.}
    \label{fig:javaPySimions}
\end{figure}

\subsection{RQ3: Across Programming Languages}
\label{sect:ra3}

We execute \approach{} on the Java and Python projects from GCJ. From 36,403 extracted snippets, \approach{} identified 131 Java and 48 Python functions clustered into 34 cross-language  clusters (single-language clusters are omitted from the RQ3 analysis). 
On validation, we find that 2 of these 34 ($5.8\%$) clusters are false positives which is better than the percentage of false positives found in Java and Python independently. That said, \approach{} would produce more clusters when support for the languages is broadened. 

We discover 616 type-III clusters by comparing the ASTs of Java and Python snippets (\tref{type3Validity}), of which 25 clusters are valid (4.1\% precision). It should be noted that this is a conservative precision estimate; the baseline was created by starting with close behavioral matches, hence giving the AST analysis a slight edge on precision (\sect{AST}).

An example of a pair of Java-Python clones can be seen in \figref{javaPySimions}. $func\_3b0e$ is a Java function that uses a loop to find the minimum in an array while $func\_6437$ is a Python function uses the inbuilt $min$ function in Python. 

\begin{tcolorbox}
\textbf{RQ3:} \approach{} succeeds in identifying clones between programming languages irrespective of their typing.
\end{tcolorbox}

\section{Discussion}
\label{sect:discuss}

We have demonstrated how \approach{} can successfully identify clones in single-language, multi-language, static typed language, and dynamic typed language environments. Compared to prior art (\baseline{}), \approach{} identifies a superset of the clusters and with higher precision. Compared to type-III clone detection, \approach{} achieves a much higher precision in Python and in cross-language situations. 
This would lead us to believe that traditional methods that detect syntactic type-III clones cannot be used for cross-language clone detection, despite successful applications in single languages for identifying libraries with reusable code~\cite{burd2002evaluating}, detecting malicious code~\cite{walenstein2007software}, catching plagiarism~\cite{baker1995finding} and identifying opportunities for refactoring~\cite{milea2014scalable}. 

Next, we explore the sensitivity of code clones to the number of inputs, the number of arguments, and the size of the snippets. 

\subsection{Impact of input sizes}
\label{sect:inputVariance}
\begin{table}[tb]
    \centering
     \caption{
    Mean and variance (in parenthesis) of \#~clones, \#~clusters and \#~false positives for 20 repeats when \#~inputs varying between 8-256. The mean (and variance) are reported.
    }
    \begin{tabular}{cccc}
        \toprule
        \textbf{\# Inputs} & \textbf{\# Clones} & \textbf{\# Clusters} & \textbf{\# False Positives}\\ \midrule
        8 & 4461(85) & 218(16) & 184(19)\\
        16  & 4297(49)& 355(17) & 142(19)\\
        32  & 4221(23) & 412(13) & 101(5)\\
        64  & 4194(4) & 623(6) & 71(3)\\
        128 & 4180(0) & 630(1) & 52(0)\\ 
        256 & 4180(0) & 632(0) & 50(0)\\ \bottomrule
    \end{tabular}
   
    \label{tab:inputVarianceCoverage}
\end{table}

Prior studies have shown that varying the number of inputs can alter the accuracy of clone detection techniques~\cite{jiang2009automatic,kim2011mecc,white2016deep}. This was particularly evident in the earliest clone detection techniques by Jiang and Su~\cite{jiang2009automatic} where the authors limited the number of inputs to 10 with a maximum of 120 permutations of the input due to the need for large computational resources and the corresponding runtime.

We test the impact on clones, clusters, and false positives by varying the number of inputs from 8 to 256 in powers of 2 and repeating \approach{} using the generated Java functions. 
Each experiment is repeated 20 times on a set of randomly generated inputs. 
For each set of input, we record the mean and variance for the number of clones, clusters and false positives, as shown in \tref{inputVarianceCoverage}. 
For a given number of inputs, each row represents the mean and variance (in parenthesis) of the number of clones, clusters and false positives. 
For low numbers of inputs, we see more functions being marked as clones and fewer clusters. 
As the number of inputs increases, the number of clones reduces and the number of clusters increases, demonstrating that the additional inputs are critical at differentiating behavior between functions. 
The counts of clones, clusters, and false positives appear to plateau after 64 inputs. 
This highlights that 10 inputs used by Jiang and Su would not be sufficient for optimally identifying true functional clones and will lead to a large number of false positives, as suggested in prior work~\cite{deissenboeck2012challenges}.

\subsection{Influence of arguments in clones}
\label{section:argmax}
\begin{figure}[tb]
    \centering
    \includegraphics[width=0.9\linewidth]{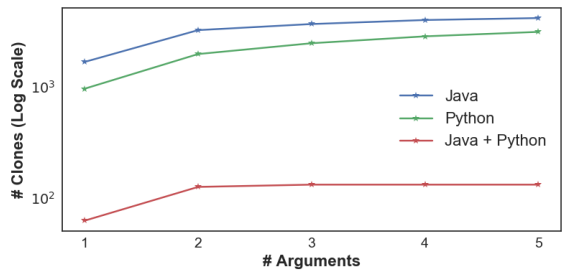}
    \caption{Cumulative \# clones with \# arguments varying between 1-5.}
    \label{fig:clonesVsArgs}
\end{figure}

We use our engineering judgment to set \mintinline{Java}{ARGS_MAX = 5} (Maximum number of Arguments) to limit the number of functions generated from snippets. \figref{clonesVsArgs} represents the cumulative number of clones with arguments varying from 1 to 5 and can be used to justify our choice of \mintinline{Java}{ARGS_MAX}. Most clones detected by \approach{} have two arguments or less. In Java functions, 3252 of 4180 clones detected have less than three arguments. 
Cross-language functions are fewer in number and typically contain functions with 2 arguments or less (125 out of 131). 
This would seem intuitive as modular functions are more frequent compared to complex functionalities. 
As \mintinline{Java}{ARGS_MAX} increases, it begins to plateau around 3. 
Hence, a larger value of \mintinline{Java}{ARGS_MAX} may not yield significantly larger number of code clones but would incur more computational resources (\mintinline{Java}{ARGS_MAX!} function executions).

\subsection{Clones vs Lines Of Code}

\begin{figure}[tb]
    \centering
    \includegraphics[width=0.9\linewidth]{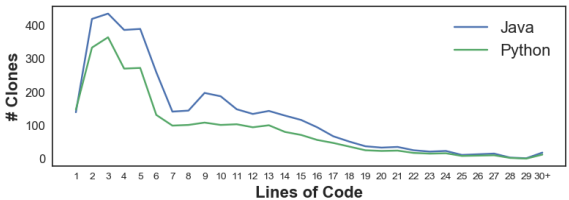}
    \caption{\# clones for lines of code between ranging from 1-29. Clones with 30 or more lines are grouped into 30+}
    \label{fig:clonesVsLoc}
\end{figure}

Prior work suggests there is more code redundancy at smaller levels of granularity~\cite{su2016identifying}. 
Aggregating all the cloned functions identified by \approach{} in RQ1, RQ2, and RQ3, we have 6,536 total, valid  cloned functions in Java and Python (duplicates removed, as the same function could be included in an RQ1 and an RQ3 cluster, for example). 

\figref{clonesVsLoc} represents the number of clones with lines of code varying from 1 to 29. Clones with 30 or more lines are denoted as ``30+". 
More than 50\% of the valid Java clones have 6 lines of code or less (2037/3845), while the median of valid Python clones have 5 lines or less (1372/2691). This implies that snippets with more lines of code are more unique and harder to clone functionally. On the contrary, smaller snippets are more likely to contain 
clones
in a code base. The greater median for Java clones compared to Python clones can be attributed to the verbosity in Java compared to the succinct nature of Python~\cite{gupta2004good}.

\section{Related Work}
\label{sect:related}

In keeping with the survey on code clones by Roy \etal~\cite{roy2009comparison}, research on code clones can broadly be classified  as \emph{syntactic}~\cite{kamiya2002ccfinder, li2004cp, baxter1998clone, jiang2007deckard, gabel2008scalable, li2012cbcd}, which represent structural similarities, and \emph{semantic}~\cite{jiang2009automatic, su2016code, su2016identifying, elva2012jsctracker}, which  represent behavioral similarities. 

EQMiner~\cite{jiang2009automatic} is the closest related work with respect to our methodology. They examined the Linux Kernel v2.6.24 by using a similar segmentation procedure, used 10 randomly generated inputs to execute them, and cluster based on IO behavior. Compared with \approach{}, EQMiner crucially ignores cross-language clone detection. Furthermore, the implementation of EQMiner contains several limitations, noted by Deissenboeck~\cite{deissenboeck2012challenges}, that make cross-language detection infeasible and even replication itself impractical. As a result, we build on the ideas pioneered by EQMiner, while overcoming limitations in its original design. We introduce novel contributions, such as using grey-box analysis to overcome the limitations of simple random random testing, scale the input generation phase from 10 to 256 inputs, which drastically reduces false positives, introduce several steps and components to support complex language features, such as lambda functions, and handle differences arising from cross-language types. Finally, \approach{} introduces flexibility in clustering as it permits a tolerance on similarity due to the \mintinline{Java}{SIM_T} hyper-parameter.

\baseline{}~\cite{su2016identifying} by Su \etal also performs simion-based comparisons to identify clones. It uses existing workloads like test-cases or `main' function calls to collect values for the behavior \added{rather than the random testing approach proposed in EQMiner or the grey-box analysis approach used in \approach{}. Research shows that existing unit tests do not attain complete code coverage~\cite{gopinath2014code} and as a result, the application of such a technique to open source repositories might not be produce a comprehensive set of clones. This conjecture can be observed in RQ1 where \approach{} identifies more clones to \baseline{} by an order of magnitude}.
Further, \baseline{} operates at a method level granularity while \approach{} can operate at method or statement level granularity. Naturally, this ensures a greater number of code clones since \approach{} can identify succinct behavior in complex code snippets.

\added{LASSO~\cite{kessel2019efficacy} by Kessel and Atkinson, like \baseline{}, is another clone detection technique for method level clones from large repositories using test cases. But unlike \baseline{}, it does not use predefined test cases; LASSO generates test cases using random generation via Evosuite~\cite{fraser2011evosuite}. That said, LASSO has many deviations compared to \baseline{} and \approach{}. Firstly, LASSO identifies only clones that have the same signature and method name (excluding case). Secondly, it detects clones only in methods where the arguments are primitive datatypes, boxed wrappers of primitives, strings, and one dimensional arrays of these datatypes. It fails to support objects;  \approach{} supports objects that can be initialized recursively using constructors of its members(\sect{args}). Finally, LASSO supports only strongly typed languages as it does not have a type inference engine like \approach{} does.}

\added{Most clone detection techniques~\cite{kamiya2002ccfinder, li2004cp, baxter1998clone, jiang2007deckard, gabel2008scalable, li2012cbcd} have been proposed for single language clone detection. With respect to cross language clone detection, we failed to find any techniques based on semantic behavior of code. A small number of techniques have been proposed on syntactic code features~\cite{nguyen2017exploring,nafi2019CLCDSA}. API2Vec~\cite{nguyen2017exploring} detects clones between two syntactically similar languages by embedding source code into a vector representation and subsequently comparing the similarity between vectors to identify code clones. CLCDSA~\cite{nafi2019CLCDSA}, identifies nine features from the source code AST and uses a deep neural network based model to learn the features and detect cross language clones.}

Segmentation used in \approach{} is inspired by methods that parse ASTs of the source code~\cite{baxter1998clone, jiang2007deckard}.  These methods encode the ASTs into intermediate representations and do not account for the semantic relationships. For example, DECKARD~\cite{jiang2007deckard} characterizes sub-trees of the AST into numerical vectors and clusters them based on the Euclidean distance which fails to capture the behavior of code in the clusters~\cite{jiang2009automatic}. This limitation has been observed in other syntactic methods as well and is a reason for adoption of semantic techniques to detect code clones~\cite{jiang2009automatic}.

\section{Limitations and Threats}
\label{sect:threats}

Threats to external validity include the focus on two languages as instances of static and dynamic typing, so results may not generalize beyond Java and Python. The use of GCJ code may not generalize to more complex code bases. 
Threats to internal validity include that for RQ3, where we ``help" the AST matching by starting with behavioral clusters and then determining if the ASTs are similar; 
which overestimates the precision of cross-language AST matching.\\
\noindent Our implementation of \approach{} has the following limitations:

\noindent \textbf{Dynamic Typing}. \approach{} does not support two primitive types \mintinline{Python}{long} and \mintinline{Python}{complex} for Python. That being said, we verified that the GCJ projects used in this study, do not explicitly use these values in the source code and they are not present in the input file used by the baseline HitoshiIO. Further, in case of a failure to identify the type of a function argument, the function was fuzzed with arguments of all supported types. In this study, we supported primitive types and the simple data-structures  \mintinline{Python}{tuple}, \mintinline{Python}{set}, \mintinline{Python}{list} and \mintinline{Python}{dict}. Support for other sophisticated data-structures can be  incorporated by extending the existing \approach{} API with instructions in the wiki~\cite{slaccTool}.

\noindent \textbf{Unsupported Features}. Although \approach{} supports Object Oriented features such as inheritance and encapsulation, it is limited to objects derived from primitive types. Hence, the current version of \approach{} cannot scale to more sophisticated objects like Threads and Database Connections. Similarly, for Python we do not support modules like generators and decorators. Nevertheless, it would be possible to support these features with more engineering effort.

\noindent \textbf{Dead Code Elimination}: In the code-clone examples of \figref{pySimions} and \figref{javaPySimions}, we  see the presence of lines of code that do not influence the return value i.e., Dead Code. At the moment, the functions do not fail due to dead code but eliminating them would make the functions more succinct and comprehensible. This will be an avenue for future work  for specific applications of \approach{}.

\section{Conclusion}
In this paper, we present \approach{}, a technique for language-agnostic code clone detection that precisely yields semantic code clones across programming languages. This is the first research to identify semantic code clones in a dynamic typed language and also across differently-typed programming languages. \approach{} identifies clones by comparing the IO relationship of segmented snippets of code from a target repository. Input values for the segmented code are generated using multi-modal grey-box fuzzing. This results in fewer false positives compared to current state of the art semantic code clone detection tool, \baseline{}.  In our study, we identify code clones between Java and Python from Google Code Jam submissions. Compared to \baseline{}, \approach{} identifies significantly ($6x$) more code clones, with greater precision ($86.7\%$ vs. $30.7\%$). 
\approach{} also detects code clones in a multi-language code corpora. The number of clones detected was fewer and the number of false positives was slightly more compared to code clones within the same language. However, future work that broadens language support is likely to improve these metrics. 
These results have implications for future applications of behavioral code clones, such as enabling robust language migration tools or mastery of a new programming language once one is known.  

\approach{} is open-source and the data used in this study is publicly available~\cite{slaccTool}.

\begin{acks}
Special thanks to Fang-Hsiang Su, Jonathan Bell, Gail Kaiser and Simha Sethumadhavan for making \baseline{} publicly available. We would also like to thank the anonymous reviewers for their valuable feedback. 
This material is based upon work supported by the National Science Foundation under Grant No. \href{https://www.nsf.gov/awardsearch/showAward?AWD_ID=1645136}{1645136} and Grant No. \href{https://www.nsf.gov/awardsearch/showAward?AWD_ID=1749936}{1749936}. 
\end{acks}

\bibliographystyle{plain}

\bibliography{arxiv_SLACC.bbl}


\begin{thebibliography}{52}


\ifx \showCODEN    \undefined \def \showCODEN     #1{\unskip}     \fi
\ifx \showDOI      \undefined \def \showDOI       #1{#1}\fi
\ifx \showISBNx    \undefined \def \showISBNx     #1{\unskip}     \fi
\ifx \showISBNxiii \undefined \def \showISBNxiii  #1{\unskip}     \fi
\ifx \showISSN     \undefined \def \showISSN      #1{\unskip}     \fi
\ifx \showLCCN     \undefined \def \showLCCN      #1{\unskip}     \fi
\ifx \shownote     \undefined \def \shownote      #1{#1}          \fi
\ifx \showarticletitle \undefined \def \showarticletitle #1{#1}   \fi
\ifx \showURL      \undefined \def \showURL       {\relax}        \fi
\providecommand\bibfield[2]{#2}
\providecommand\bibinfo[2]{#2}
\providecommand\natexlab[1]{#1}
\providecommand\showeprint[2][]{arXiv:#2}

\bibitem[\protect\citeauthoryear{Baker}{Baker}{1995}]%
        {baker1995finding}
\bibfield{author}{\bibinfo{person}{Brenda~S Baker}.}
  \bibinfo{year}{1995}\natexlab{}.
\newblock \showarticletitle{On finding duplication and near-duplication in
  large software systems}. In \bibinfo{booktitle}{\emph{Proceedings of 2nd
  Working Conference on Reverse Engineering}}. IEEE, \bibinfo{pages}{86--95}.
\newblock


\bibitem[\protect\citeauthoryear{Barr, Brun, Devanbu, Harman, and Sarro}{Barr
  et~al\mbox{.}}{2014}]%
        {barr2014plastic}
\bibfield{author}{\bibinfo{person}{Earl~T Barr}, \bibinfo{person}{Yuriy Brun},
  \bibinfo{person}{Premkumar Devanbu}, \bibinfo{person}{Mark Harman}, {and}
  \bibinfo{person}{Federica Sarro}.} \bibinfo{year}{2014}\natexlab{}.
\newblock \showarticletitle{The plastic surgery hypothesis}. In
  \bibinfo{booktitle}{\emph{Proceedings of the 22nd ACM SIGSOFT International
  Symposium on Foundations of Software Engineering}}. ACM,
  \bibinfo{pages}{306--317}.
\newblock


\bibitem[\protect\citeauthoryear{Baxter, Yahin, Moura, Sant'Anna, and
  Bier}{Baxter et~al\mbox{.}}{1998}]%
        {baxter1998clone}
\bibfield{author}{\bibinfo{person}{Ira~D Baxter}, \bibinfo{person}{Andrew
  Yahin}, \bibinfo{person}{Leonardo Moura}, \bibinfo{person}{Marcelo
  Sant'Anna}, {and} \bibinfo{person}{Lorraine Bier}.}
  \bibinfo{year}{1998}\natexlab{}.
\newblock \showarticletitle{Clone detection using abstract syntax trees}. In
  \bibinfo{booktitle}{\emph{Software Maintenance, 1998. Proceedings.,
  International Conference on}}. IEEE, \bibinfo{pages}{368--377}.
\newblock


\bibitem[\protect\citeauthoryear{Beit-Aharon}{Beit-Aharon}{2002}]%
        {beit2002source}
\bibfield{author}{\bibinfo{person}{Jonathan Beit-Aharon}.}
  \bibinfo{year}{2002}\natexlab{}.
\newblock \bibinfo{title}{Source code translation}.
\newblock
\newblock
\newblock
\shownote{US Patent App. 15/894,096.}


\bibitem[\protect\citeauthoryear{Bowles and Bethke~Jr}{Bowles and
  Bethke~Jr}{1983}]%
        {bowles1983multi}
\bibfield{author}{\bibinfo{person}{Stephen~W Bowles} {and}
  \bibinfo{person}{George~E Bethke~Jr}.} \bibinfo{year}{1983}\natexlab{}.
\newblock \bibinfo{title}{Multi-pass system and method for source to source
  code translation}.
\newblock
\newblock
\newblock
\shownote{US Patent 4,374,408.}


\bibitem[\protect\citeauthoryear{Burd and Bailey}{Burd and Bailey}{2002}]%
        {burd2002evaluating}
\bibfield{author}{\bibinfo{person}{Elizabeth Burd} {and} \bibinfo{person}{John
  Bailey}.} \bibinfo{year}{2002}\natexlab{}.
\newblock \showarticletitle{Evaluating clone detection tools for use during
  preventative maintenance}. In \bibinfo{booktitle}{\emph{Proceedings. Second
  IEEE International Workshop on Source Code Analysis and Manipulation}}. IEEE,
  \bibinfo{pages}{36--43}.
\newblock


\bibitem[\protect\citeauthoryear{Cousot and Cousot}{Cousot and Cousot}{1977}]%
        {cousot1977abstract}
\bibfield{author}{\bibinfo{person}{Patrick Cousot} {and}
  \bibinfo{person}{Radhia Cousot}.} \bibinfo{year}{1977}\natexlab{}.
\newblock \showarticletitle{Abstract interpretation: a unified lattice model
  for static analysis of programs by construction or approximation of
  fixpoints}. In \bibinfo{booktitle}{\emph{Proceedings of the 4th ACM
  SIGACT-SIGPLAN symposium on Principles of programming languages}}. ACM,
  \bibinfo{pages}{238--252}.
\newblock


\bibitem[\protect\citeauthoryear{Dang, Zhang, Ge, Huang, Chu, and Xie}{Dang
  et~al\mbox{.}}{2017}]%
        {Dang:2017}
\bibfield{author}{\bibinfo{person}{Yingnong Dang}, \bibinfo{person}{Dongmei
  Zhang}, \bibinfo{person}{Song Ge}, \bibinfo{person}{Ray Huang},
  \bibinfo{person}{Chengyun Chu}, {and} \bibinfo{person}{Tao Xie}.}
  \bibinfo{year}{2017}\natexlab{}.
\newblock \showarticletitle{Transferring Code-clone Detection and Analysis to
  Practice}. In \bibinfo{booktitle}{\emph{Proceedings of the 39th International
  Conference on Software Engineering: Software Engineering in Practice Track}}
  (Buenos Aires, Argentina) \emph{(\bibinfo{series}{ICSE-SEIP '17})}.
  \bibinfo{publisher}{IEEE Press}, \bibinfo{address}{Piscataway, NJ, USA},
  \bibinfo{pages}{53--62}.
\newblock
\showISBNx{978-1-5386-2717-4}
\urldef\tempurl%
\url{https://doi.org/10.1109/ICSE-SEIP.2017.6}
\showDOI{\tempurl}


\bibitem[\protect\citeauthoryear{Deissenboeck, Heinemann, Hummel, and
  Wagner}{Deissenboeck et~al\mbox{.}}{2012}]%
        {deissenboeck2012challenges}
\bibfield{author}{\bibinfo{person}{Florian Deissenboeck}, \bibinfo{person}{Lars
  Heinemann}, \bibinfo{person}{Benjamin Hummel}, {and} \bibinfo{person}{Stefan
  Wagner}.} \bibinfo{year}{2012}\natexlab{}.
\newblock \showarticletitle{Challenges of the dynamic detection of functionally
  similar code fragments}. In \bibinfo{booktitle}{\emph{Software Maintenance
  and Reengineering (CSMR), 2012 16th European Conference on}}. IEEE,
  \bibinfo{pages}{299--308}.
\newblock


\bibitem[\protect\citeauthoryear{Elva and Leavens}{Elva and Leavens}{2012}]%
        {elva2012jsctracker}
\bibfield{author}{\bibinfo{person}{Rochelle Elva} {and} \bibinfo{person}{Gary~T
  Leavens}.} \bibinfo{year}{2012}\natexlab{}.
\newblock \bibinfo{booktitle}{\emph{Jsctracker: A semantic clone detection tool
  for java code}}.
\newblock \bibinfo{type}{{T}echnical {R}eport}.
  \bibinfo{institution}{University of Central Florida, Dept. of EECS, CS
  division}.
\newblock


\bibitem[\protect\citeauthoryear{Fau and Bihler}{Fau and Bihler}{[n.d.]}]%
        {java2Csharp}
\bibfield{author}{\bibinfo{person}{Alexandre Fau} {and}
  \bibinfo{person}{Reinhold Bihler}.} \bibinfo{year}{[n.d.]}\natexlab{}.
\newblock \bibinfo{title}{Java2CSharp}.
\newblock
  \bibinfo{howpublished}{\url{http://sourceforge.net/projects/j2cstranslator/}}.
\newblock
\newblock
\shownote{Accessed: 2018-09-25.}


\bibitem[\protect\citeauthoryear{Fjeldberg}{Fjeldberg}{2008}]%
        {fjeldberg2008:polyglot}
\bibfield{author}{\bibinfo{person}{Hans-Christian Fjeldberg}.}
  \bibinfo{year}{2008}\natexlab{}.
\newblock \emph{\bibinfo{title}{Polyglot programming}}.
\newblock \bibinfo{thesistype}{Ph.D. Dissertation}. \bibinfo{school}{Master
  thesis, Norwegian University of Science and Technology, Trondheim/Norway}.
\newblock


\bibitem[\protect\citeauthoryear{Fraser and Arcuri}{Fraser and Arcuri}{2011}]%
        {fraser2011evosuite}
\bibfield{author}{\bibinfo{person}{Gordon Fraser} {and} \bibinfo{person}{Andrea
  Arcuri}.} \bibinfo{year}{2011}\natexlab{}.
\newblock \showarticletitle{EvoSuite: automatic test suite generation for
  object-oriented software}. In \bibinfo{booktitle}{\emph{Proceedings of the
  19th ACM SIGSOFT symposium and the 13th European conference on Foundations of
  software engineering}}. ACM, \bibinfo{pages}{416--419}.
\newblock


\bibitem[\protect\citeauthoryear{Gabel, Jiang, and Su}{Gabel
  et~al\mbox{.}}{2008}]%
        {gabel2008scalable}
\bibfield{author}{\bibinfo{person}{Mark Gabel}, \bibinfo{person}{Lingxiao
  Jiang}, {and} \bibinfo{person}{Zhendong Su}.}
  \bibinfo{year}{2008}\natexlab{}.
\newblock \showarticletitle{Scalable detection of semantic clones}. In
  \bibinfo{booktitle}{\emph{Proceedings of the 30th international conference on
  Software engineering}}. ACM, \bibinfo{pages}{321--330}.
\newblock


\bibitem[\protect\citeauthoryear{Google}{Google}{[n.d.]}]%
        {googleCodeJam}
\bibfield{author}{\bibinfo{person}{Google}.} \bibinfo{year}{[n.d.]}\natexlab{}.
\newblock \bibinfo{title}{Google Code Jam}.
\newblock \bibinfo{howpublished}{\url{code.google.com/codejam}}.
\newblock
\newblock
\shownote{Accessed: 2018-09-25.}


\bibitem[\protect\citeauthoryear{Gopinath, Jensen, and Groce}{Gopinath
  et~al\mbox{.}}{2014}]%
        {gopinath2014code}
\bibfield{author}{\bibinfo{person}{Rahul Gopinath}, \bibinfo{person}{Carlos
  Jensen}, {and} \bibinfo{person}{Alex Groce}.}
  \bibinfo{year}{2014}\natexlab{}.
\newblock \showarticletitle{Code coverage for suite evaluation by developers}.
  In \bibinfo{booktitle}{\emph{Proceedings of the 36th International Conference
  on Software Engineering}}. ACM, \bibinfo{pages}{72--82}.
\newblock


\bibitem[\protect\citeauthoryear{Gupta}{Gupta}{2004}]%
        {gupta2004good}
\bibfield{author}{\bibinfo{person}{Diwaker Gupta}.}
  \bibinfo{year}{2004}\natexlab{}.
\newblock \showarticletitle{What is a good first programming language?}
\newblock \bibinfo{journal}{\emph{Crossroads}} \bibinfo{volume}{10},
  \bibinfo{number}{4} (\bibinfo{year}{2004}), \bibinfo{pages}{7--7}.
\newblock


\bibitem[\protect\citeauthoryear{Jensen, M{\o}ller, and Thiemann}{Jensen
  et~al\mbox{.}}{2009}]%
        {jensen2009type}
\bibfield{author}{\bibinfo{person}{Simon~Holm Jensen}, \bibinfo{person}{Anders
  M{\o}ller}, {and} \bibinfo{person}{Peter Thiemann}.}
  \bibinfo{year}{2009}\natexlab{}.
\newblock \showarticletitle{Type analysis for JavaScript}. In
  \bibinfo{booktitle}{\emph{International Static Analysis Symposium}}.
  Springer, \bibinfo{pages}{238--255}.
\newblock


\bibitem[\protect\citeauthoryear{Jiang, Misherghi, Su, and Glondu}{Jiang
  et~al\mbox{.}}{2007}]%
        {jiang2007deckard}
\bibfield{author}{\bibinfo{person}{Lingxiao Jiang}, \bibinfo{person}{Ghassan
  Misherghi}, \bibinfo{person}{Zhendong Su}, {and} \bibinfo{person}{Stephane
  Glondu}.} \bibinfo{year}{2007}\natexlab{}.
\newblock \showarticletitle{Deckard: Scalable and accurate tree-based detection
  of code clones}. In \bibinfo{booktitle}{\emph{Proceedings of the 29th
  international conference on Software Engineering}}. IEEE Computer Society,
  \bibinfo{pages}{96--105}.
\newblock


\bibitem[\protect\citeauthoryear{Jiang and Su}{Jiang and Su}{2009}]%
        {jiang2009automatic}
\bibfield{author}{\bibinfo{person}{Lingxiao Jiang} {and}
  \bibinfo{person}{Zhendong Su}.} \bibinfo{year}{2009}\natexlab{}.
\newblock \showarticletitle{Automatic mining of functionally equivalent code
  fragments via random testing}. In \bibinfo{booktitle}{\emph{Proceedings of
  the eighteenth international symposium on Software testing and analysis}}.
  ACM, \bibinfo{pages}{81--92}.
\newblock


\bibitem[\protect\citeauthoryear{Kamiya, Kusumoto, and Inoue}{Kamiya
  et~al\mbox{.}}{2002}]%
        {kamiya2002ccfinder}
\bibfield{author}{\bibinfo{person}{Toshihiro Kamiya}, \bibinfo{person}{Shinji
  Kusumoto}, {and} \bibinfo{person}{Katsuro Inoue}.}
  \bibinfo{year}{2002}\natexlab{}.
\newblock \showarticletitle{CCFinder: a multilinguistic token-based code clone
  detection system for large scale source code}.
\newblock \bibinfo{journal}{\emph{IEEE Transactions on Software Engineering}}
  \bibinfo{volume}{28}, \bibinfo{number}{7} (\bibinfo{year}{2002}),
  \bibinfo{pages}{654--670}.
\newblock


\bibitem[\protect\citeauthoryear{Kessel and Atkinson}{Kessel and
  Atkinson}{2019}]%
        {kessel2019efficacy}
\bibfield{author}{\bibinfo{person}{Marcus Kessel} {and} \bibinfo{person}{Colin
  Atkinson}.} \bibinfo{year}{2019}\natexlab{}.
\newblock \showarticletitle{On the Efficacy of Dynamic Behavior Comparison for
  Judging Functional Equivalence}. In \bibinfo{booktitle}{\emph{2019 19th
  International Working Conference on Source Code Analysis and Manipulation
  (SCAM)}}. IEEE, \bibinfo{pages}{193--203}.
\newblock


\bibitem[\protect\citeauthoryear{Khan, Khan, et~al\mbox{.}}{Khan
  et~al\mbox{.}}{2012}]%
        {khan2012comparative}
\bibfield{author}{\bibinfo{person}{Mohd~Ehmer Khan}, \bibinfo{person}{Farmeena
  Khan}, {et~al\mbox{.}}} \bibinfo{year}{2012}\natexlab{}.
\newblock \showarticletitle{A comparative study of white box, black box and
  grey box testing techniques}.
\newblock \bibinfo{journal}{\emph{Int. J. Adv. Comput. Sci. Appl}}
  \bibinfo{volume}{3}, \bibinfo{number}{6} (\bibinfo{year}{2012}).
\newblock


\bibitem[\protect\citeauthoryear{Kim, Jung, Kim, and Yi}{Kim
  et~al\mbox{.}}{2011}]%
        {kim2011mecc}
\bibfield{author}{\bibinfo{person}{Heejung Kim}, \bibinfo{person}{Yungbum
  Jung}, \bibinfo{person}{Sunghun Kim}, {and} \bibinfo{person}{Kwankeun Yi}.}
  \bibinfo{year}{2011}\natexlab{}.
\newblock \showarticletitle{MeCC: memory comparison-based clone detector}. In
  \bibinfo{booktitle}{\emph{Proceedings of the 33rd International Conference on
  Software Engineering}}. ACM, \bibinfo{pages}{301--310}.
\newblock


\bibitem[\protect\citeauthoryear{Koschke, Falke, and Frenzel}{Koschke
  et~al\mbox{.}}{2006}]%
        {koschke2006clone}
\bibfield{author}{\bibinfo{person}{Rainer Koschke}, \bibinfo{person}{Raimar
  Falke}, {and} \bibinfo{person}{Pierre Frenzel}.}
  \bibinfo{year}{2006}\natexlab{}.
\newblock \showarticletitle{Clone detection using abstract syntax suffix
  trees}. In \bibinfo{booktitle}{\emph{2006 13th Working Conference on Reverse
  Engineering}}. IEEE, \bibinfo{pages}{253--262}.
\newblock


\bibitem[\protect\citeauthoryear{Li and Ernst}{Li and Ernst}{2012}]%
        {li2012cbcd}
\bibfield{author}{\bibinfo{person}{Jingyue Li} {and} \bibinfo{person}{Michael~D
  Ernst}.} \bibinfo{year}{2012}\natexlab{}.
\newblock \showarticletitle{CBCD: Cloned buggy code detector}. In
  \bibinfo{booktitle}{\emph{Proceedings of the 34th International Conference on
  Software Engineering}}. IEEE Press, \bibinfo{pages}{310--320}.
\newblock


\bibitem[\protect\citeauthoryear{Li, Lu, Myagmar, and Zhou}{Li
  et~al\mbox{.}}{2004}]%
        {li2004cp}
\bibfield{author}{\bibinfo{person}{Zhenmin Li}, \bibinfo{person}{Shan Lu},
  \bibinfo{person}{Suvda Myagmar}, {and} \bibinfo{person}{Yuanyuan Zhou}.}
  \bibinfo{year}{2004}\natexlab{}.
\newblock \showarticletitle{CP-Miner: A Tool for Finding Copy-paste and Related
  Bugs in Operating System Code.}. In \bibinfo{booktitle}{\emph{OSdi}},
  Vol.~\bibinfo{volume}{4}. \bibinfo{pages}{289--302}.
\newblock


\bibitem[\protect\citeauthoryear{Mathew, Parnin, and Stolee}{Mathew
  et~al\mbox{.}}{[n.d.]}]%
        {slaccTool}
\bibfield{author}{\bibinfo{person}{George Mathew}, \bibinfo{person}{Chris
  Parnin}, {and} \bibinfo{person}{Kathryn~T Stolee}.}
  \bibinfo{year}{[n.d.]}\natexlab{}.
\newblock \bibinfo{title}{SLACC}.
\newblock
  \bibinfo{howpublished}{\url{github.com/DynamicCodeSearch/SLACC/tree/ICSE20}}.
\newblock
\newblock
\shownote{[Online; accessed 06-February-2020].}


\bibitem[\protect\citeauthoryear{Mayer and Bauer}{Mayer and Bauer}{2015}]%
        {Mayer:2015}
\bibfield{author}{\bibinfo{person}{Philip Mayer} {and}
  \bibinfo{person}{Alexander Bauer}.} \bibinfo{year}{2015}\natexlab{}.
\newblock \showarticletitle{An Empirical Analysis of the Utilization of
  Multiple Programming Languages in Open Source Projects}. In
  \bibinfo{booktitle}{\emph{Proceedings of the 19th International Conference on
  Evaluation and Assessment in Software Engineering}} (Nanjing, China)
  \emph{(\bibinfo{series}{EASE '15})}. \bibinfo{publisher}{ACM},
  \bibinfo{address}{New York, NY, USA}, Article \bibinfo{articleno}{4},
  \bibinfo{numpages}{10}~pages.
\newblock
\showISBNx{978-1-4503-3350-4}
\urldef\tempurl%
\url{https://doi.org/10.1145/2745802.2745805}
\showDOI{\tempurl}


\bibitem[\protect\citeauthoryear{Mayer, Kirsch, and Le}{Mayer
  et~al\mbox{.}}{2017}]%
        {Mayer2017}
\bibfield{author}{\bibinfo{person}{Philip Mayer}, \bibinfo{person}{Michael
  Kirsch}, {and} \bibinfo{person}{Minh~Anh Le}.}
  \bibinfo{year}{2017}\natexlab{}.
\newblock \showarticletitle{On multi-language software development,
  cross-language links and accompanying tools: a survey of professional
  software developers}.
\newblock \bibinfo{journal}{\emph{Journal of Software Engineering Research and
  Development}} \bibinfo{volume}{5}, \bibinfo{number}{1} (\bibinfo{date}{19
  Apr} \bibinfo{year}{2017}), \bibinfo{pages}{1}.
\newblock
\showISSN{2195-1721}
\urldef\tempurl%
\url{https://doi.org/10.1186/s40411-017-0035-z}
\showDOI{\tempurl}


\bibitem[\protect\citeauthoryear{Milea, Jiang, and Khoo}{Milea
  et~al\mbox{.}}{2014}]%
        {milea2014scalable}
\bibfield{author}{\bibinfo{person}{Narcisa~Andreea Milea},
  \bibinfo{person}{Lingxiao Jiang}, {and} \bibinfo{person}{Siau-Cheng Khoo}.}
  \bibinfo{year}{2014}\natexlab{}.
\newblock \showarticletitle{Scalable detection of missed cross-function
  refactorings}. In \bibinfo{booktitle}{\emph{Proceedings of the 2014
  International Symposium on Software Testing and Analysis}}. ACM,
  \bibinfo{pages}{138--148}.
\newblock


\bibitem[\protect\citeauthoryear{Nafi, Sheka~Kar, Roy, K.~Roy, and
  Schneider}{Nafi et~al\mbox{.}}{[n.d.]}]%
        {nafi2019CLCDSA}
\bibfield{author}{\bibinfo{person}{Kawser Nafi}, \bibinfo{person}{Tonny
  Sheka~Kar}, \bibinfo{person}{Banani Roy}, \bibinfo{person}{Chanchal K.~Roy},
  {and} \bibinfo{person}{Kevin Schneider}.} \bibinfo{year}{[n.d.]}\natexlab{}.
\newblock \showarticletitle{CLCDSA: Cross Language Code Clone Detection using
  Syntactical Features and API Documentation}.
\newblock  (\bibinfo{year}{[n.\,d.]}).
\newblock


\bibitem[\protect\citeauthoryear{Nguyen, Nguyen, Phan, and Nguyen}{Nguyen
  et~al\mbox{.}}{2017}]%
        {nguyen2017exploring}
\bibfield{author}{\bibinfo{person}{Trong~Duc Nguyen}, \bibinfo{person}{Anh~Tuan
  Nguyen}, \bibinfo{person}{Hung~Dang Phan}, {and} \bibinfo{person}{Tien~N
  Nguyen}.} \bibinfo{year}{2017}\natexlab{}.
\newblock \showarticletitle{Exploring API embedding for API usages and
  applications}. In \bibinfo{booktitle}{\emph{Software Engineering (ICSE), 2017
  IEEE/ACM 39th International Conference on}}. IEEE, \bibinfo{pages}{438--449}.
\newblock


\bibitem[\protect\citeauthoryear{{Python Community}}{{Python
  Community}}{[n.d.]}]%
        {pythonAST}
\bibfield{author}{\bibinfo{person}{{Python Community}}.}
  \bibinfo{year}{[n.d.]}\natexlab{}.
\newblock \bibinfo{title}{Python AST}.
\newblock \bibinfo{howpublished}{\url{docs.python.org/3/library/ast.html}}.
\newblock
\newblock
\shownote{[Online; accessed 23-August-2019].}


\bibitem[\protect\citeauthoryear{Ray, Kim, Person, and Rungta}{Ray
  et~al\mbox{.}}{2013}]%
        {Ray:2013}
\bibfield{author}{\bibinfo{person}{Baishakhi Ray}, \bibinfo{person}{Miryung
  Kim}, \bibinfo{person}{Suzette Person}, {and} \bibinfo{person}{Neha Rungta}.}
  \bibinfo{year}{2013}\natexlab{}.
\newblock \showarticletitle{Detecting and Characterizing Semantic
  Inconsistencies in Ported Code}. In \bibinfo{booktitle}{\emph{Proceedings of
  the 28th IEEE/ACM International Conference on Automated Software
  Engineering}} (Silicon Valley, CA, USA) \emph{(\bibinfo{series}{ASE'13})}.
  \bibinfo{publisher}{IEEE Press}, \bibinfo{address}{Piscataway, NJ, USA},
  \bibinfo{pages}{367--377}.
\newblock
\showISBNx{978-1-4799-0215-6}
\urldef\tempurl%
\url{https://doi.org/10.1109/ASE.2013.6693095}
\showDOI{\tempurl}


\bibitem[\protect\citeauthoryear{Roy, Cordy, and Koschke}{Roy
  et~al\mbox{.}}{2009}]%
        {roy2009comparison}
\bibfield{author}{\bibinfo{person}{Chanchal~K Roy}, \bibinfo{person}{James~R
  Cordy}, {and} \bibinfo{person}{Rainer Koschke}.}
  \bibinfo{year}{2009}\natexlab{}.
\newblock \showarticletitle{Comparison and evaluation of code clone detection
  techniques and tools: A qualitative approach}.
\newblock \bibinfo{journal}{\emph{Science of computer programming}}
  \bibinfo{volume}{74}, \bibinfo{number}{7} (\bibinfo{year}{2009}),
  \bibinfo{pages}{470--495}.
\newblock


\bibitem[\protect\citeauthoryear{Scholtz and Wiedenbeck}{Scholtz and
  Wiedenbeck}{1990}]%
        {Scholtz:subsequent}
\bibfield{author}{\bibinfo{person}{Jean Scholtz} {and} \bibinfo{person}{Susan
  Wiedenbeck}.} \bibinfo{year}{1990}\natexlab{}.
\newblock \showarticletitle{Learning second and subsequent programming
  languages: A problem of transfer}.
\newblock \bibinfo{journal}{\emph{International Journal of Human-Computer
  Interaction}} \bibinfo{volume}{2}, \bibinfo{number}{1}
  (\bibinfo{year}{1990}), \bibinfo{pages}{51--72}.
\newblock
\urldef\tempurl%
\url{https://doi.org/10.1080/10447319009525970}
\showDOI{\tempurl}


\bibitem[\protect\citeauthoryear{{Shrestha}, {Barik}, and {Parnin}}{{Shrestha}
  et~al\mbox{.}}{2018}]%
        {Shrestha}
\bibfield{author}{\bibinfo{person}{N. {Shrestha}}, \bibinfo{person}{T.
  {Barik}}, {and} \bibinfo{person}{C. {Parnin}}.}
  \bibinfo{year}{2018}\natexlab{}.
\newblock \showarticletitle{{It's Like Python But: Towards Supporting Transfer
  of Programming Language Knowledge}}. In \bibinfo{booktitle}{\emph{2018 IEEE
  Symposium on Visual Languages and Human-Centric Computing (VL/HCC)}}.
  \bibinfo{pages}{177--185}.
\newblock
\showISSN{1943-6106}
\urldef\tempurl%
\url{https://doi.org/10.1109/VLHCC.2018.8506508}
\showDOI{\tempurl}


\bibitem[\protect\citeauthoryear{Su, Bell, Harvey, Sethumadhavan, Kaiser, and
  Jebara}{Su et~al\mbox{.}}{2016a}]%
        {su2016code}
\bibfield{author}{\bibinfo{person}{Fang-Hsiang Su}, \bibinfo{person}{Jonathan
  Bell}, \bibinfo{person}{Kenneth Harvey}, \bibinfo{person}{Simha
  Sethumadhavan}, \bibinfo{person}{Gail Kaiser}, {and} \bibinfo{person}{Tony
  Jebara}.} \bibinfo{year}{2016}\natexlab{a}.
\newblock \showarticletitle{Code relatives: detecting similarly behaving
  software}. In \bibinfo{booktitle}{\emph{Proceedings of the 2016 24th ACM
  SIGSOFT International Symposium on Foundations of Software Engineering}}.
  ACM, \bibinfo{pages}{702--714}.
\newblock


\bibitem[\protect\citeauthoryear{Su, Bell, Kaiser, and Sethumadhavan}{Su
  et~al\mbox{.}}{2016b}]%
        {su2016identifying}
\bibfield{author}{\bibinfo{person}{Fang-Hsiang Su}, \bibinfo{person}{Jonathan
  Bell}, \bibinfo{person}{Gail Kaiser}, {and} \bibinfo{person}{Simha
  Sethumadhavan}.} \bibinfo{year}{2016}\natexlab{b}.
\newblock \showarticletitle{Identifying functionally similar code in complex
  codebases}. In \bibinfo{booktitle}{\emph{Program Comprehension (ICPC), 2016
  IEEE 24th International Conference on}}. IEEE, \bibinfo{pages}{1--10}.
\newblock


\bibitem[\protect\citeauthoryear{{Team GitHub}}{{Team GitHub}}{[n.d.]}]%
        {gists}
\bibfield{author}{\bibinfo{person}{{Team GitHub}}.}
  \bibinfo{year}{[n.d.]}\natexlab{}.
\newblock \bibinfo{title}{GitHub Gist}.
\newblock \bibinfo{howpublished}{\url{https://gist.github.com/discover}}.
\newblock
\newblock
\shownote{[Online; accessed 23-August-2019].}


\bibitem[\protect\citeauthoryear{{Team Stack Overflow}}{{Team Stack
  Overflow}}{[n.d.]}]%
        {stackoverflow}
\bibfield{author}{\bibinfo{person}{{Team Stack Overflow}}.}
  \bibinfo{year}{[n.d.]}\natexlab{}.
\newblock \bibinfo{title}{Stack Overflow}.
\newblock \bibinfo{howpublished}{\url{https://stackoverflow.com}}.
\newblock
\newblock
\shownote{[Online; accessed 23-August-2019].}


\bibitem[\protect\citeauthoryear{Tomassetti and Torchiano}{Tomassetti and
  Torchiano}{2014}]%
        {Tomassetti:2014}
\bibfield{author}{\bibinfo{person}{Federico Tomassetti} {and}
  \bibinfo{person}{Marco Torchiano}.} \bibinfo{year}{2014}\natexlab{}.
\newblock \showarticletitle{An Empirical Assessment of Polyglot-ism in GitHub}.
  In \bibinfo{booktitle}{\emph{Proceedings of the 18th International Conference
  on Evaluation and Assessment in Software Engineering}} (London, England,
  United Kingdom) \emph{(\bibinfo{series}{EASE '14})}.
  \bibinfo{publisher}{ACM}, \bibinfo{address}{New York, NY, USA}, Article
  \bibinfo{articleno}{17}, \bibinfo{numpages}{4}~pages.
\newblock
\showISBNx{978-1-4503-2476-2}
\urldef\tempurl%
\url{https://doi.org/10.1145/2601248.2601269}
\showDOI{\tempurl}


\bibitem[\protect\citeauthoryear{van Bruggen}{van Bruggen}{2015}]%
        {javaParser}
\bibfield{author}{\bibinfo{person}{Danny van Bruggen}.}
  \bibinfo{year}{2015}\natexlab{}.
\newblock \bibinfo{title}{Javaparser - For processing Java code}.
\newblock \bibinfo{howpublished}{\url{github.com/javaparser/javaparser}}.
\newblock
\newblock
\shownote{[Online; accessed 23-August-2019].}


\bibitem[\protect\citeauthoryear{Walenstein and Lakhotia}{Walenstein and
  Lakhotia}{2007}]%
        {walenstein2007software}
\bibfield{author}{\bibinfo{person}{Andrew Walenstein} {and}
  \bibinfo{person}{Arun Lakhotia}.} \bibinfo{year}{2007}\natexlab{}.
\newblock \showarticletitle{The software similarity problem in malware
  analysis}. In \bibinfo{booktitle}{\emph{Dagstuhl Seminar Proceedings}}.
  Schloss Dagstuhl-Leibniz-Zentrum f{\"u}r Informatik.
\newblock


\bibitem[\protect\citeauthoryear{White, Tufano, Vendome, and Poshyvanyk}{White
  et~al\mbox{.}}{2016}]%
        {white2016deep}
\bibfield{author}{\bibinfo{person}{Martin White}, \bibinfo{person}{Michele
  Tufano}, \bibinfo{person}{Christopher Vendome}, {and} \bibinfo{person}{Denys
  Poshyvanyk}.} \bibinfo{year}{2016}\natexlab{}.
\newblock \showarticletitle{Deep learning code fragments for code clone
  detection}. In \bibinfo{booktitle}{\emph{Proceedings of the 31st IEEE/ACM
  International Conference on Automated Software Engineering}}. ACM,
  \bibinfo{pages}{87--98}.
\newblock


\bibitem[\protect\citeauthoryear{{Wikipedia Contributors}}{{Wikipedia
  Contributors}}{[n.d.]}]%
        {javaByteCode}
\bibfield{author}{\bibinfo{person}{{Wikipedia Contributors}}.}
  \bibinfo{year}{[n.d.]}\natexlab{}.
\newblock \bibinfo{title}{Java bytecode instruction listings}.
\newblock \bibinfo{howpublished}{\url{en.wikipedia.org/wiki/Java_bytecode}}.
\newblock
\newblock
\shownote{[Online; accessed 23-August-2019].}


\bibitem[\protect\citeauthoryear{{Wikipedia contributors}}{{Wikipedia
  contributors}}{2019}]%
        {wiki:levenshtein}
\bibfield{author}{\bibinfo{person}{{Wikipedia contributors}}.}
  \bibinfo{year}{2019}\natexlab{}.
\newblock \bibinfo{title}{Levenshtein distance --- {Wikipedia}{,} The Free
  Encyclopedia}.
\newblock
  \bibinfo{howpublished}{\url{en.wikipedia.org/wiki/Levenshtein_distance}}.
\newblock
\newblock
\shownote{[Online; accessed 23-August-2019].}


\bibitem[\protect\citeauthoryear{Wu and Anderson}{Wu and Anderson}{1990}]%
        {quanfeng}
\bibfield{author}{\bibinfo{person}{Quanfeng Wu} {and} \bibinfo{person}{John~R.
  Anderson}.} \bibinfo{year}{1990}\natexlab{}.
\newblock \bibinfo{booktitle}{\emph{{Problem-solving transfer among programming
  languages}}}.
\newblock \bibinfo{type}{{T}echnical {R}eport}. \bibinfo{institution}{Carnegie
  Mellon University}.
\newblock


\bibitem[\protect\citeauthoryear{Wyrich, Graziotin, and Wagner}{Wyrich
  et~al\mbox{.}}{2019}]%
        {10.7717/peerj-cs.173}
\bibfield{author}{\bibinfo{person}{Marvin Wyrich}, \bibinfo{person}{Daniel
  Graziotin}, {and} \bibinfo{person}{Stefan Wagner}.}
  \bibinfo{year}{2019}\natexlab{}.
\newblock \showarticletitle{A theory on individual characteristics of
  successful coding challenge solvers}.
\newblock \bibinfo{journal}{\emph{PeerJ Computer Science}}  \bibinfo{volume}{5}
  (\bibinfo{date}{Feb.} \bibinfo{year}{2019}), \bibinfo{pages}{e173}.
\newblock
\showISSN{2376-5992}
\urldef\tempurl%
\url{https://doi.org/10.7717/peerj-cs.173}
\showDOI{\tempurl}


\bibitem[\protect\citeauthoryear{Yang}{Yang}{1991}]%
        {yang1991identifying}
\bibfield{author}{\bibinfo{person}{Wuu Yang}.} \bibinfo{year}{1991}\natexlab{}.
\newblock \showarticletitle{Identifying syntactic differences between two
  programs}.
\newblock \bibinfo{journal}{\emph{Software: Practice and Experience}}
  \bibinfo{volume}{21}, \bibinfo{number}{7} (\bibinfo{year}{1991}),
  \bibinfo{pages}{739--755}.
\newblock


\bibitem[\protect\citeauthoryear{{Yue}, {Gao}, {Meng}, {Xiong}, {Wang}, and
  {Morgenthaler}}{{Yue} et~al\mbox{.}}{2018}]%
        {Yue:2018}
\bibfield{author}{\bibinfo{person}{R. {Yue}}, \bibinfo{person}{Z. {Gao}},
  \bibinfo{person}{N. {Meng}}, \bibinfo{person}{Y. {Xiong}},
  \bibinfo{person}{X. {Wang}}, {and} \bibinfo{person}{J.~D. {Morgenthaler}}.}
  \bibinfo{year}{2018}\natexlab{}.
\newblock \showarticletitle{Automatic Clone Recommendation for Refactoring
  Based on the Present and the Past}. In \bibinfo{booktitle}{\emph{2018 IEEE
  International Conference on Software Maintenance and Evolution (ICSME)}}.
  \bibinfo{pages}{115--126}.
\newblock
\showISSN{2576-3148}
\urldef\tempurl%
\url{https://doi.org/10.1109/ICSME.2018.00021}
\showDOI{\tempurl}


\end{thebibliography}

\balance

\end{document}